\documentclass[prb,aps,10pt,amsmath,amssymb,floatfix,twocolumn,superscriptaddress]{revtex4-1}

\usepackage[english]{babel}
\usepackage[utf8x]{inputenc}
\usepackage{graphicx}
\usepackage{amsmath}
\usepackage{amssymb}
\usepackage{nicefrac}
\usepackage{hyperref}
\usepackage{xspace}

\let\AAt\AA
\renewcommand{\AA}{\ifmmode{\mathrm{\AAt}}\else{\AAt}\fi\xspace}

\newcommand{\unit}[1]{\ensuremath{\mathrm{#1}}\xspace}

\newcommand{\chem}[1]{\ensuremath{\rm #1}\xspace}
\newcommand{\mB}{{\rm \mu_B}}
\newcommand{\hole}{\ensuremath{\mathrm{\mathbf{h}}}\xspace}

\newcommand{\iv}{I--V\xspace}

\preprint{Gate Spin Control v2.1}

\begin{document}

	\title{Gate Control of Spin-Polarized Conductance\\in Alloyed Transition Metal Nano-contacts}
	
	\author{Ilia~N.~Sivkov}
	\affiliation{ETH Z\"urich / CSCS Lugano, Switzerland}
	\affiliation{Max Planck Institute of Microstructure Physics, Halle, Germany}
	\author{Oleg~O.~Brovko}
	\affiliation{The Abdus Salam International Centre for Theoretical Physics (ICTP), Trieste, Italy}
	\author{Ivan Rungger}
	\affiliation{National Physical Laboratory, Teddington TW11 0LW, UK}
	\author{Valeri~S.~Stepanyuk}
	\affiliation{Max Planck Institute of Microstructure Physics, Halle, Germany}
	
	\begin{abstract}
		To date, endeavors in nanoscale spintronics are dominated by the use of single-electron or single-spin transistors having at their heart a semiconductor, metallic or molecular quantum dot who's localized states are non-spin-degenerate and can be controlled by an external bias applied via a gate electrode. Adjusting the bias of the gate one can realign those states with respect to the chemical potentials of the leads and thus tailor the spin-polarized transmission properties of the device. Here we show that similar functionality can be achieved in a purely metallic junction comprised of a metallic magnetic chains attached to metallic paramagnetic leads and biased by a gate electrode. Our \textit{ab initio} calculations of electron transport through mixed Pt-Fe (Fe-Pd and Fe-Rh) atomic chains suspended between Pt (Pd and Rh) electrodes show that spin-polarized confined states of the chain can be shifted by the gate bias causing a change in the relative contributions of majority and minority channels to the nano-contact's conductance. As a result, we observe strong dependence of conductance spin-polarization on the applied gate potential. In some cases the spin-polarization of conductance can even be reversed in sign upon gate potential application, which is a remarkable and promising trait for spintronic applications.
	\end{abstract}

	\maketitle
	
\section*{Introduction}
	
	As spintronics evolves from a holy grail of science and technology into an everyday phenomenon, its foundations demand for the same cornerstones as conventional spin-degenerate electronics is built on: miniaturization, robustness and tunability. Technological progress nowadays has enabled both scientists and engineers to build nanometer and sub-nanometer-scale devices not only with elaborate and volatile break junction,~\cite{Krans1993,Champagne2005,Hauptmann2008,Egle2010} electromigration~\cite{Park1999} or STM extraction~\cite{Brandbyge1995} techniques, but also with virtually integration-ready nanotube/nanowire and lithography based technologies.~\cite{Bernand-Mantel2006,Franklin2012,Parekh2012,Lin2013,Pribiag2013,Desai2016} Robustness of transport properties is usually achieved by a combination of single-electron and coulomb-blockade physics~\cite{Averin1986,Schelp1997,Takahashi1998,Likharev1999,Bernand-Mantel2006} with strong magnetic properties guaranteed by the use of magnetic materials with high spin moment~\cite{Bernand-Mantel2006,Rocha2007} and geometry/topology based stability of the latter.~\cite{Pontes2008,Smogunov2008,Kim2012} The tunability, or the susceptibility of spin-dependent transport properties~\cite{Tsymbal2003,Requist2016} to external perturbations can be achieved in several ways. Orientation and ordering of spins in magnetic systems can be ``programmed'' into atomic-scale junctions at construction stage (\textit{e.g.}, by tuning the chemical composition and geometry of the device \cite{Bernand-Mantel2006,Rocha2007,Smogunov2008,Pontes2008,Tao2009,Egle2010,Tsukamoto2013,Sivkov2014a}), but it is even more important and challenging to be able to dynamically and reversibly tune those properties after the junction has been completed or integrated into a circuit. The ``traditional'' way of controlling transmission properties of both conventional and single-electron transistors is the field effect, \textit{i.e.} external bias application by means of an additional \emph{gate} electrode. Gate bias in a three-terminal device can be used to control both paramagnetic transport~\cite{Champagne2005,Obermair2010,Osorio2008} and magnetic characteristics of a nanoscale junction, such as its magnetoresistive or spin-filtering properties.~\cite{Hauptmann2008,Pribiag2013,Lee2013,Modepalli2016,Wang2016}
	
	Due to basic physical and technological reasons, the above-listed criteria for a spintronic nanojunction are usually realized by using, as the heart of the junction, semiconductor quantum dots or nanowires,~\cite{Pribiag2013,Modepalli2016} metallic quantum dots decoupled from the rest of the system by vacuum or insulating layer~\cite{Bernand-Mantel2006} or nanotubes and molecules.~\cite{Hauptmann2008,Lee2013,Wang2016} The chemical composition and geometry of the junction then guarantee the single-electron physics and the spin-selectiveness, while a bias applied to the system via a gate electrode is used to manipulate the local electronic structure of the quantum dot, nanotube or molecule thus changing its magnetic properties.
	
	In the present paper we show that the same functionality of a spin-filtering junction with gate-controlled magnetic properties can be achieved in purely metallic magnetic nano-contacts making use of quantum well states confined to the junction and the effect of electric field screening. Guided by the criteria listed above, we take as a sample system alloyed magnetic $5d-3d$ nanowires connected to paramagnetic electrodes. Using short chains suspended between compact electrodes one achieves minimal device dimensions. The use of magnetic $3d$ transition-metal atoms as a source of high spin moment in conjunction with high anisotropy provided by the strong spin-orbit coupling of the $5d$ elements~\cite{Smogunov2008,Dasa2012} guarantees pronounced and robust static and transport magnetic properties of the device. Similar properties have been found in conducting molecular junctions, where magnetic selectiveness is achieved due to the system's geometry/symmetry.~\cite{Smogunov2015,Li2016} Finally, electric field bias has been repeatedly shown to have a strong effect on the magnetism of metallic nanostructures.~\cite{Dasa2012,Brovko2014}
	
	As a particular example we show, by \textit{ab initio} calculations, that spin-polarized transport through alloyed $3d-4d$ (Pd-Fe, Rh-Fe) and $3d-5d$ (Pt-Fe) metallic chains suspended between metallic leads (respectively Pd or Pt) has a strong spin-polarized character and can be deliberately and reversibly tuned by a gate bias in a wide range of values. In some cases the gate bias is shown to even lead to an inversion in conductance spin-polarization (SP).
	
\section*{Methods and Geometries}
	All calculations of transport properties were performed with a density functional theory (DFT) based code \textit{Smeagol}.~\cite{Rocha2005,Rocha2006,Rungger2008} implementing the Keldysh formalism for non-equilibrium Green's functions (NEGF). For the geometric relaxation of the system geometries a DFT code Siesta was used.~\cite{Ordejon1996,Soler2002} For the investigation of electronic effects related to the influence of a gate bias, we used functionality available in the \textit{Smeagol} code.
	
\begin{figure}
		\center{\includegraphics{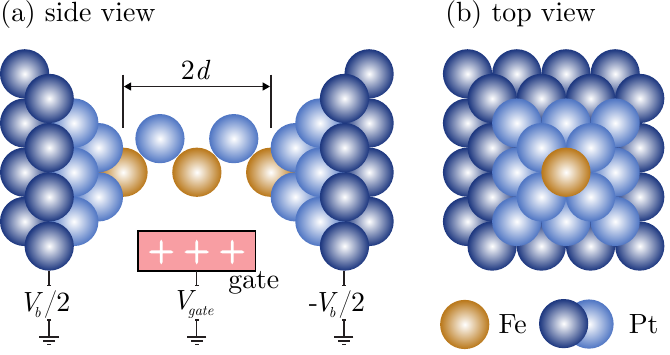}}
		\caption{Schematic view of the nano-contact: alloyed $3d-5d$ transition metal chain suspended between two electrodes with a gate electrode placed abreast the center of the chain. The length of the junction is parameterized by the distance $2d$ between the electrodes' apex atoms.}
		\label{fig:geom}
	\end{figure}

	The studied nano-contacts were modeled as mixed Pt-Fe (Pd-Fe, Rh-Fe) 5-atomic zigzag/linear chains suspended between Pt (Pd, Rh) electrodes as sketched in Fig.~\ref{fig:geom}. Electrodes comprised 15-atom pyramids joined to planar leads consisting of 8 monolayers each with an in-plane unit cell of $4 \times 4$ atoms per layer. Before the final construction of the nano-contact the electrodes were fully relaxed. To investigate transport properties of different possible structural configurations of the nano-contacts we performed calculations for different stretching states of the nano-contact chain, parameterized by the distance $d$ between the Fe atoms (the interelectrode distance in Fig.~\ref{fig:geom} thus being $2d$). Since lattice constants of bulk Pt,Pd and Rh differ, we used different ranges for $d$-values to keep the nano-contact chain in a stable state. For Pt-Fe, Pd-Fe and Rh-Fe the ranges of $d$ were chosen to be $2.6-4.75\AA$, $2.4-4.6\AA$ and $3.6-4.6\AA$, respectively, which covers different stretching stages from compressed or zigzag to linear.
	
	In both geometry optimizazation and transport calculations a $k$-mesh of $5 \times 5$ Monkhorst-Pack~\cite{Monkhorst1976} distributed points was used. In the following, whenever we presen the results of spin-resolved calculation and refer to majority and minority (or spin-up and spin-down) electrons, we use as a reference the local electronic structure of the central Fe atom of the junction.
	
	The presence of a gate electrode was modeled by adding a constant background charge to a box-shaped volume next to the nano-contact. The electrostatic potential, created by the charged gate was added to the Hartree potential which was then used in the DFT-NEGF scheme. In our case the box had dimensions of $4.0 \times 1.5 \times 4.6\AA$ and resided some 2.8\AA away from the axis of the contact, as shown in Fig.~\ref{fig:geom}. For the calculations we used charges of $q=+1$, $0$ and $-1\hole$, where $\hole=-e$ is the charge of an electron hole. The electrostatic gate potential felt by the chain ranged from $\sim 0.5~\unit{V}$ around the central Fe atom to almost $0~\unit{V}$ at the leads (see Fig.~\ref{fig:pot_chredist}a for a gate potential map). As our further analysis has shown, electron and transport properties practically do not depend on the polarity of the electric field, but only on its magnitude. This is, of course, not a universal property of a gated junction, but applies to our chosen system in the most relevant cases of the chain being either fully stretched or close to linear. For strongly buckled chains the polarity of the gate bias shall be determinant to the bond polarization and screening patterns and thus to the effect of the gate on the electronic, magnetic and transport properties of the junction. While it might be argued that placing an electrode $2.8\AA$ from the junction is an unjustifiably complex task to achieve experimentally, the conclusions of the paper shall not change if the electrode is placed further away from the junction and the potential of the gate is increased to compensate for the increased distance. An alternative could be to immerse the nano-contact into ionic liquid thus in effect amplifying the bias.~\cite{Weisheit2007} Albeit, the presence of an ionic solution in the junction would most likely either destroy the wire or the action of the ionic adlayer would drastically and unpredictably change the magnetic configuration of the nanowire through charge doping or direct chemical bonding.
	
	
\section*{Results and Discussion}
	
	To set the stage for the main results we have to shortly discuss the ground state magnetic properties of the investigated systems. As our calculations show, the spin arrangement of Fe atoms is robust against stretching/geometry alteration of the nano-contact, the ground state for all geometries (with a sole exception of the Pt-Fe contact with $d=2.6\AA$) being an antiferromagnetic (AF) one. The AF order of Fe spins is stabilized by the super-exchange interaction via the Pt (Pd, Rh) atoms interspacing the latter. Since the electrodes are non-magnetic inherently (Pt, Pd and Rh can acquire a small magnetic moment induced by Fe proximity), all spin-dependent interactions occur in the contact chain and all observed spin-dependent properties should be related to the chain's electronic and geometric structures.

	\begin{figure}
		\center\includegraphics{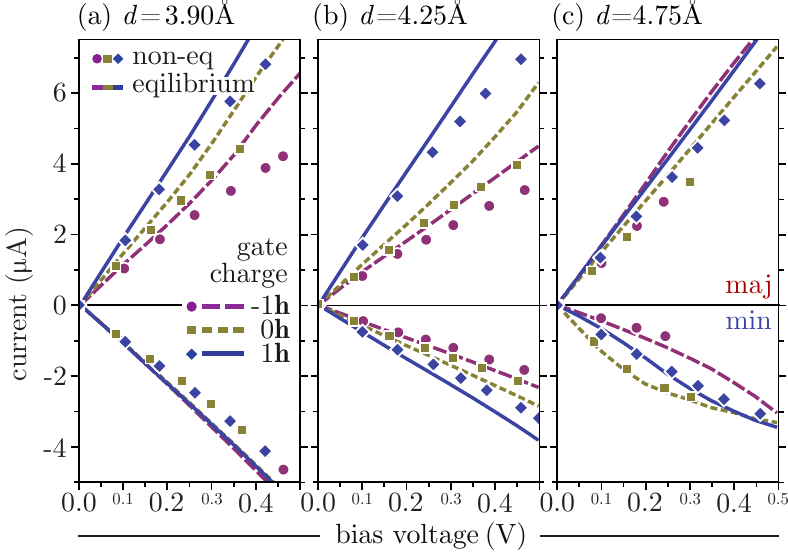}
		\caption{Equilibrium (lines) and non-equilibrium (symbols) spin-polarized \iv characteristics of a Pt-Fe nano-contact at different inter-electrode distances ($d=3.90$ (a), $4.25$ (b) and $4.75\AA$ (c)) for gate electrode biases of $-1$, $0$ and $1\hole$ (purple solid line and circles, green dashed line and squares and blue dotted lines and rhombs, respectively). The majority \iv curves are plotted above the abscissa and the minority ones below.}
		\label{fig:iv}
	\end{figure}
	
\begin{figure}
		\center\includegraphics{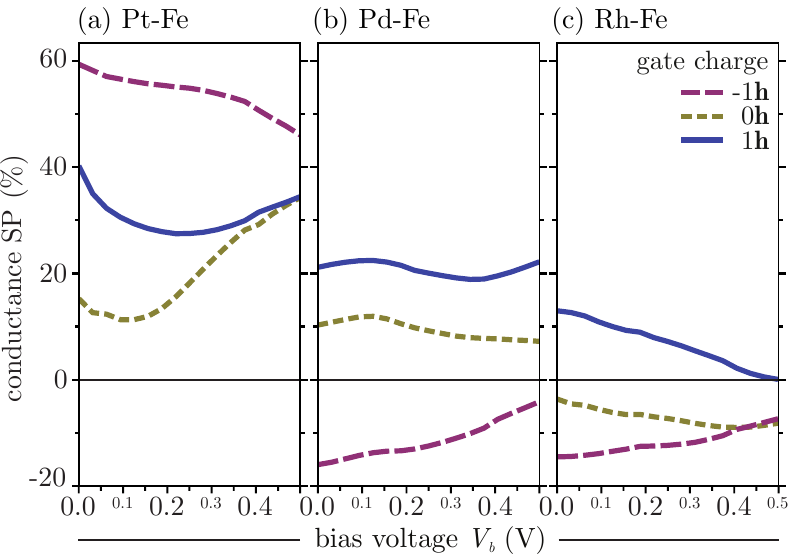}
		\caption{Spin-polarization of conductance through Pt-Fe (a), Rh-Fe (b) and Pd-Fe (c) nano-contacts at gate biases of -1,0 and 1\hole (purple solid, green dashed and blue dotted lines respectively) as a function of the inter-electrode bias $V_b$. The corresponding inter-lead distances $d$ were $4.75$, $4.6$ and $4.0\AA$, respectively.}
		\label{fig:ivsp}
	\end{figure}

	The studied junctions exhibit conductive properties typical for metallic systems. Taking Fe-Pt junction as a representative example (Fig.~\ref{fig:iv}, the other studied systems exhibited qualitatively identical behavior) we calculate both non-equilibrium and equilibrium spin-resolved current-voltage (\iv) characteristics. The equilibrium estimates are done by integrating the zero-bias transmission around the Fermi level which, as the results show, is a fair approximation to the fully self-consistent non-equilibrium transport at small biases ($0.1-0.3~\unit{V}$) when the electronic structure of the leads is not strongly changed by the applied bias. We find nearly linear trends in both spin channels for different stretching states ($d$) of the junction (the stretching range covers the transition from the zig-zag to linear configurations, as discussed in the supplement). There are several important implications this results has for our study. Firstly, we note the spin-polarized nature of the electron transport as the slopes of the \iv curves are distinctly different in different spin channels. Secondly, as opposed to single-channel systems, the junction exhibits metallic-like behavior (nigh-linear behavior of the \iv curves) at all inter-electrode distances. A careful inspection, however, shows that the \iv characteristics of the nano-contact are not precisely linear but exhibit slight deviations from a strict Ohm's law. Those deviations are not equally pronounced in the two spin channels and are strongest for the stretched junction (linear configuration) hinting at the importance of both electronic properties and geometry of the contact for the control over spin-polarized conductance. Fig.~\ref{fig:ivsp}a shows (as a green dashed curve) the bias voltage ($V_b$) resolved value of the SP of conductance through the stretched Pt-Fe nano-contact exhibiting a change of the SP from 20\% to 40\% with the voltage increase. At smaller distances Pt atoms form a parallel non-spin-polarized conduction channel somewhat reducing the effect.
	
	But the most important feature of the \iv curves in Fig.~\ref{fig:iv} is the dependence of the the latter on the applied gate potential. While the green dashed line and squares in Fig.~\ref{fig:iv} show the \iv characteristics for a system unbiased by the gate electrode, the purple solid line and circles and blue dotted line and rhombs show the \iv curves for the gate being charged to $-1\hole$ or $+1\hole$ respectively. For shorter contact chains the effect of the bias lies in a change of the \iv curve slope, but for a stretched contact, where the \iv curve was exhibiting a non-linearity even in the unbiased case, the gate charge causes a significant change in the system's conducting behavior. This change is of a different character and of different order of magnitude in majority and minority spin channels, which brings about a change of zero-bias conductance SP from 60\% to 15\% to 40\% (at $-1$, $0$ and $+1\hole$ gate charges respectively). This remarkable trend shows that one can use gating to voluntarily alter the SP of electron transport through a nanojunction as evidenced by Fig.~\ref{fig:ivsp}a. 
	
	This behavior is not unique to the Pt-Fe junction described above, but is equally present in the other two studied chemical compositions of the nano-contact. Figs.~\ref{fig:ivsp}b and \ref{fig:ivsp}c show the SP of conductance of Pd-Fe and Rh-Fe junctions respectively in representative states of stretching ($d=4.6\AA$ for Pd-Fe and $d=4.0\AA$ for Rh-Fe). Here (in Pd and Rh based systems) the observed behavior is even more intriguing since in both those cases it is found that the gate bias can not only alter the magnitude of the SP but even reverse its sign altogether. In general, Pd an Rh based systems show more pronounced magnetic behavior for small inter-electrode distances $d$, which is caused by relaxation patterns slightly different from those observed for Pt-Fe. The difference is not surprizing since contrary to fairly ``large'' Pt atoms, Rh and Pd are closer in ionic size to Fe atoms.
	
	\begin{figure}
		\center\includegraphics{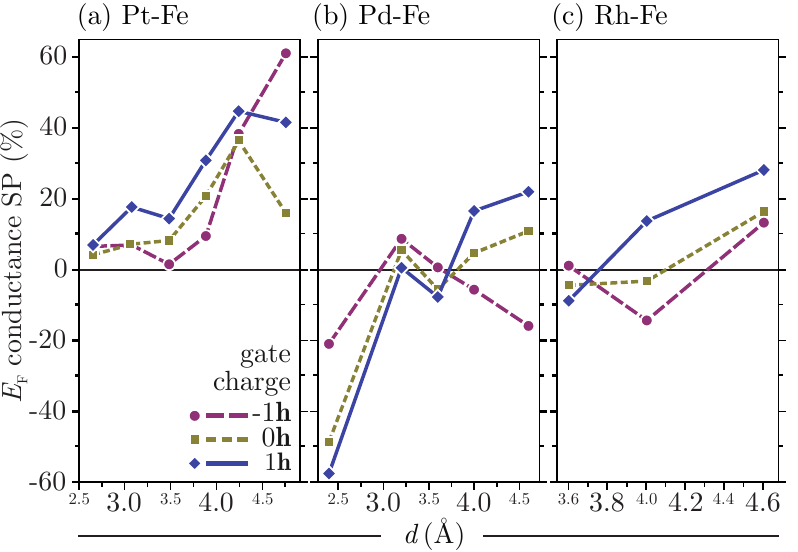}
		\caption{Summary of the dependence of conductance of the nano-contact on chemical composition (Pt-Fe, Rh-Fe and Pd-Fe in (a), (b) and (c), respectively), geometry (defined by inter-Fe distance $d$) and gate bias $V_g$ defined by the gate charge.}
		\label{fig:condsp}
	\end{figure}

	The summary of the zero-bias conductance SP for different chemical compositions and different inter-electrode distances $d$ is given in Fig.~\ref{fig:condsp}. It is clear that all three parameters (chemistry, geometry and gate bias) play a crucial role in determining the conductance SP. Changing the geometry and chemistry of the nano-contact one can thus program predefined conduction magnetic properties into the junction at construction stage and subsequently voluntarily alter them by applying a gate bias.
	
	\begin{figure}
		\center\includegraphics{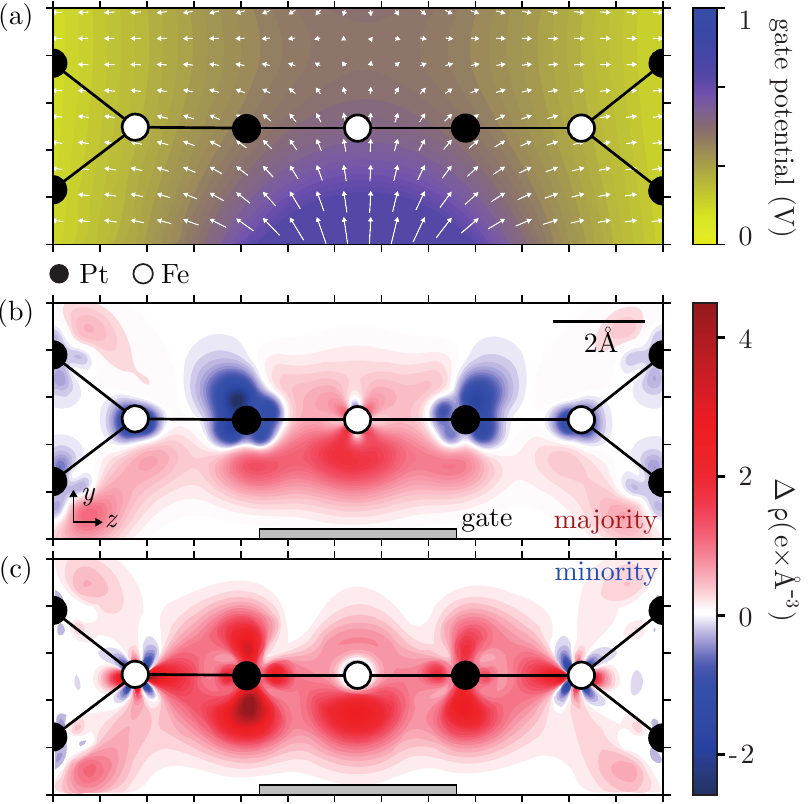}
		\caption{(a) Electrostatic potential (color-coded) cerated by a gate electrode charged to 1\hole and the ensuing electric field (arrows, length proportional to the field strength). Charge redistribution $\Delta \rho$ of majority (b) and minority (c) electrons of a linear Fe-Pt nano-junction due to the presence of a biased gate electrode charged to 1\hole. Here $\Delta \rho_\sigma = \rho_\sigma (V_{gate}=V_{1\hole}) - \rho_\sigma (V_{gate}=0)$, where $\sigma$ denotes one of the two spin channels.}
		\label{fig:pot_chredist}
	\end{figure}
	
	Let us now examine the underlying physical mechanisms leading the the remarkable susceptibility of the conductance spin-polarization to the gate bias. As a representative example, let us consider the Pt-Fe nano-contact at $d=4.75\AA$. This particular case shows most clearly the physical origins of the effect, while other systems and geometries are quite similar qualitatively. Contrary to the electronic mechanism of bias control over a single-electron quantum-dot-based junctions, where the gate bias rigidly shifts the localized levels of the quantum dot, in our metallic system the gate electrode acts more subtly on the nano-contact chain and its conduction. In particular, the presence of a charged electrode creates a varying electrostatic potential in the region of the nano-contact. Fig.~\ref{fig:pot_chredist}a shows the potential distribution and the corresponding vectors of ensuing electric field around the Pt-Fe nano-contact at a gate charge of 1\hole. The atoms of the nano-contact respond to the externally applied electric field by a charge redistribution striving to screen the perturbation. The charge displacement maps for electrons with majority and minority spin are presented in Figs.~\ref{fig:pot_chredist}b and \ref{fig:pot_chredist}c respectively. It is easy to see that the effect is strongest in the center of the nano-contact, where the external electric field is at its strongest, and gradually decays towards the leads, where the gate bias causes a weak cushion of screening charge to be formed at the surface of the contacts contributing to a change in boundary conditions for the electron transport through the alloyed magnetic chain. Yet most important for us is the apparent difference in the screening patterns of majority and minority electrons in Figs.~\ref{fig:pot_chredist}b and \ref{fig:pot_chredist}c. Different screening patterns at different atoms and in different spin channels hint that different atomic orbitals take part in the screening process. While such behavior can reasonably be expected from inherently magnetic Fe atoms, charge redistribution maps in Figs.~\ref{fig:pot_chredist}b and \ref{fig:pot_chredist}c show that Pt atoms of the chain, in fact, show a more pronounced difference in screening patterns between majority and minority channels than the neighboring Fe atoms. This behavior is not unique to the linear (stretched) geometric configuration of the nano-contact but is true for compressed, zig-zag shaped junctions as well (see Fig.~\ref{fig:s:pot_chredist}).
	
	\begin{figure}
		\center\includegraphics{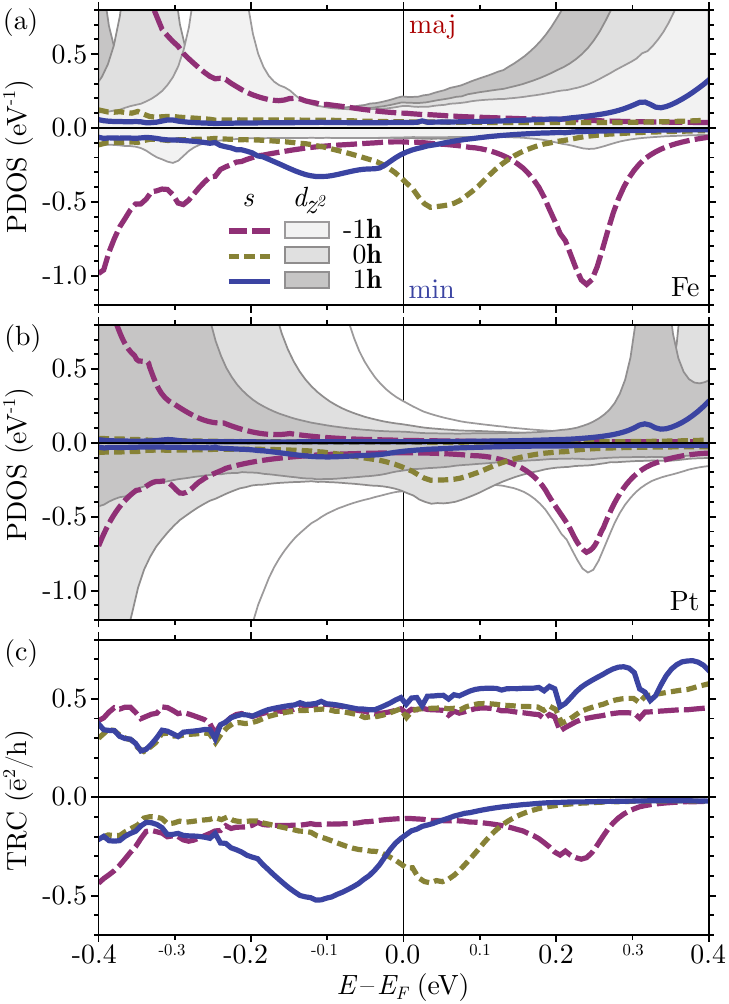}
		\caption{Spin-resolved projected density of $s$ (lines) and $d_{tot}$ (shaded areas) states for (a) central Fe and (b) neighboring Pt atoms for gate biases of $-1$, $0$ and $1\hole$. (c) Energy and spin resolved transmission probability for the Pt-Fe system with $d=4.75\AA$ for the same gate biases. The peak in minority transmission probability passing through the Fermi level is responsible for the remarkable gate-induced change in the transmission spin-polarization. Curves for majority electrons are above the abscissa and the minority curves are below.}
		\label{fig:pdos_trc}
	\end{figure}

	To see what orbitals are involved in the screening process and what ramifications the screening has for the electronic structure and the energy-resolved conductance of the nano-contact we plot in Fig.~\ref{fig:pdos_trc} the projected density of states (PDOS) and transmission probability of our test system. Figs.~\ref{fig:pdos_trc}a and \ref{fig:pdos_trc}b clearly show a sizable gate-bias-induced shift of a PDOS peak in minority $s$ states in the vicinity of the Fermi level in both Pt and Fe electronic structures. Due to $s-d_{z^2}$ hybridization a peak at corresponding energies also appears in the $d$ density of states of Pt. Since $s$ orbitals of Pt have a slow-decaying wave function tails and often play the leading role in the electron transport, a peak in minority channel transmission (Fig. \ref{fig:pdos_trc}(c)) following the $s$-peak Pt-PDOS is easy to explain. Since the zero-bias conductance is determined by the transmission coefficient at the Fermi level and \iv characteristic can be approximately seen as an integral of the transmission coefficient curve in the energy range of $E_F-V_b/2$ to $E_F+V_b/2$,~\footnote{The validity of this approximation is corroborated by the similarity in equilibrium and non-equilibrium conductance curves in Fig.~\ref{fig:iv}} the displacement of the above-mentioned transmission peaks explains the gate-induced changes in the spin polarization of the conductance and \iv characteristics.

	The existence of the peaks and their propensity to shift under the action of an external bias resembles the behavior of the localized electron level in a quantum dot, making it interesting to addressed the nature of the minority peaks in Fe and Pt PDOS which form the transmission channels. Absent in bulk and surface systems, their presence is generally attributed to electron confinement in the nano-contact's chain bound by two electrodes leading to the appearance of standing waves and localized electronic states.~\cite{Stepanyuk2004a,Emberly1999} Another way of describing the origin of the peaks (as is discussed in our previous work~\cite{Sivkov2014}) is by considering the chain as a molecular entity, in which case the peaks around the Fermi level can be seen as antibonding states of the chain living in the localizing chain potential and complementary to the bonding states lying some $1~\unit{eV}$ lower in energy (see the wide-range PDOS plots in Fig.~\ref{fig:s:pdos_wide}). Importantly for us, such states have discrete spectrum as the localized states of a quantum dot do and tend to shift monotonously with the gate bias. In the quantum well picture the shift can be seen as a result of potential well depth change acompanied with the change of boundary conditions due to the gate-induced screening charge accumulation or depletion at the leads. In the antibonding-state picture the shift corresponds to the stark-like shift of the orbital energies, a gate-bias-induced change in the bonding-antibonding splitting and the charge-redistribution-induced slight change of the magnetic moments of Fe atoms (see Fig.~\ref{fig:s:magmom}) of the nano-contact with the ensuing change in the Stoner splitting of the majority and minority $d$-states. The three contributions are hard to disentangle, yet their cumulative action can be well seen in Fig.~\ref{fig:s:pdos_wide}.

	It should be noted that the electron-confining behavior and the mechanism of gate-induced manipulation of the spin-polarization of conductance are not unique to the linear nano-contacts. Similar confined states and their reaction to the gate bias can be observed for different stretching states of the Fe-Pt nano-contact (see Fig.~\ref{fig:s:trc_dist}).
	
	The same considerations as presented above for Pt-Fe nano-contact can be applied to other systems in terms of chemical composition. For example, in Rh-Fe system the localized $s$ states are spin-split and cross the Fermi level at different gate biases, resulting the remarkable inversion of the conductance spin-polarization shown in Figs.~\ref{fig:ivsp}b and \ref{fig:condsp}b. The mechanism responsible for the switching is analogous to the one described above (see Fig.~\ref{fig:s:rhfe}). The Pd-Fe system exhibits a slightly more complex behavior, involving an interplay of many electronic states, but also resulting in the inversion of the conductance spin polarization.
	
	Even in a non-alloyed system, as, for example, a nano-contact consisting of a pure Fe chain, the gate bias shall have an impact on the electronic structure of the nano-contact and thus on the spin-polarized and paramagnetic transport properties thereof (see Fig.~\ref{fig:s:pure_fe}). In this particular case, however, the strong ferromagnetic coupling between adjacent Fe atoms create a stable ferromagnetic unit of the whole chain, giving it more robust yet less manipulatable spin-filtering transport properties.


	At this point it is also essential to remark on a few points, which call for further investigation. One such point is the issue of magnetic anisotropy. For realistic spintronic applications the relative orientation of the current spin-polarization and the nanojunction's principal magnetic axes and planes shall be pivotal to the operation of the device and it is thus essential to establish how the gate action changes the aforementioned anisotropy. Based on existing studies of the anisotropy of nanowires and its influence on electron transport,~\cite{Smogunov2008,Requist2016} our expectation is that the magnetic anisotropy in our junction shall be dictated by the geometry of the latter, i.e. the bonds and bond angles between magnetic and non-magnetic atoms of the nano-contact.~\cite{Smogunov2008,Dasa2012,Rau2014,Tsysar2012} The action of the gate may distort the geometry of the nano-contact, change the interaction between the constituent atoms thereof and by that alter the anisotropy of the whole system. If, however, the chain would be constructed as an ad-chain on an insulating substrate rather than being suspended between two electrodes in vacuum, the anusotropy shall come from the surface and be less susceptible to the action of the gate. It is, however, likely that the impact of anisotropy shall be limited to the initial electronic properties or a quantitative gauge of the gate effect. The main message of the paper shall, we believe, withstand any potential effects of the anisotropy. Another aspect definitely worth raising is the impact of many-body effects in the nano-contact. Kondo screening, along with the anisotropy, is known to affect the electron transport through a nano-contact.~\cite{Lipinski2010,Kim2012,Requist2016} How the gate bias shall impact both those phenomena is a question which falls outside the scope of the present study yet undoubtedly deserves to be answered.
	
	Another issue of critical importance is the bane of every purely theoretical study - the issue of how realistic the studied system is and how easy it is to realise experimentally. Without launching into lengthy speculations, here we just mention that alloyed metallic chains or a system with physical properties similar to those discussed in the present work can likely be constructed using advanced tip functionalisation capabilities of local probe techniques~\cite{Meyer2001} to construct the nano-junction and a multi-probe setup would provide the gating capabilities. Another way would be to use atomic manipulation to construct the junction in an atom-by-atom fashion in 2D on a wide-bandgap insulating substrate, atom manipulation being now well mastered not only on metallic,~\cite{Khajetoorians2011} but also semiconductor or insulator surfaces.~\cite{Hirjibehedin2006,Rau2014,Otte2009} This path would also presumably simplify the addition of the gate electrode to the system, which could be constructed in mesoscopic dimensions with island self-assembly or lithographic techniques.
	
	Finally, in the light of the spintronics application claims the question is bound to arise as to how the spin-polarization of conductance (controlled by the gate potential) shall be detected or further utilized. Without pretending to suggest a market ready solution we shall remark on but a few of the numerous ways of using spin-polarized current, namely by introducing another polarized barrier in the leads (by means of a spin-filtering layer~\cite{Sivkov2014a} or by sampling the electrons from the edge-state of a spin-hall lead~\cite{Brune2012}) thus creating (in optical terms) a crossed polarizer or polarizer-analyzer setup.

\section*{Conclusions}

	To summarize, we have shown by the example of mixed Pt-Fe/Pd/Rh nano-contacts shaped as mixed chains suspended between non-magnetic (Pt/Pd/Rh) electrodes that the spin-polarized conductance through the system can be controlled not just by the chemical composition and geometry tuning, but also by the application of an external potential by means of a gate electrode. Depending of the particular system, the gate potential can cause significant changes or even reversal of the conductance spin-polarization, which is a remarkable observation and a promising trait for spintronic applications. We hope that this investigation shall incite further interest in gate-control over the spin-polarized transport in atomic-scale metallic junctions.
	
\section*{Acknowledgements}

	The work was supported by Deutsche Forschungsgemeinschaft through grant SFB 762 and the project ``Structure and magnetism of cluster ensembles on metal
surfaces: Microscopic theory of the fundamental interactions''. O.B. acknowledges support from the ERC Grant No. 320796, MODPHYSFRICT. I.R. acknowledges financial support from the EU project ACMOL (FET Young Explorers, Grant No. 618082).
	

\begin{thebibliography}{53}%
	\makeatletter
	\providecommand \@ifxundefined [1]{%
	 \@ifx{#1\undefined}
	}%
	\providecommand \@ifnum [1]{%
	 \ifnum #1\expandafter \@firstoftwo
	 \else \expandafter \@secondoftwo
	 \fi
	}%
	\providecommand \@ifx [1]{%
	 \ifx #1\expandafter \@firstoftwo
	 \else \expandafter \@secondoftwo
	 \fi
	}%
	\providecommand \natexlab [1]{#1}%
	\providecommand \enquote  [1]{``#1''}%
	\providecommand \bibnamefont  [1]{#1}%
	\providecommand \bibfnamefont [1]{#1}%
	\providecommand \citenamefont [1]{#1}%
	\providecommand \href@noop [0]{\@secondoftwo}%
	\providecommand \href [0]{\begingroup \@sanitize@url \@href}%
	\providecommand \@href[1]{\@@startlink{#1}\@@href}%
	\providecommand \@@href[1]{\endgroup#1\@@endlink}%
	\providecommand \@sanitize@url [0]{\catcode `\\12\catcode `\$12\catcode
	  `\&12\catcode `\#12\catcode `\^12\catcode `\_12\catcode `\%12\relax}%
	\providecommand \@@startlink[1]{}%
	\providecommand \@@endlink[0]{}%
	\providecommand \url  [0]{\begingroup\@sanitize@url \@url }%
	\providecommand \@url [1]{\endgroup\@href {#1}{\urlprefix }}%
	\providecommand \urlprefix  [0]{URL }%
	\providecommand \Eprint [0]{\href }%
	\providecommand \doibase [0]{http://dx.doi.org/}%
	\providecommand \selectlanguage [0]{\@gobble}%
	\providecommand \bibinfo  [0]{\@secondoftwo}%
	\providecommand \bibfield  [0]{\@secondoftwo}%
	\providecommand \translation [1]{[#1]}%
	\providecommand \BibitemOpen [0]{}%
	\providecommand \bibitemStop [0]{}%
	\providecommand \bibitemNoStop [0]{.\EOS\space}%
	\providecommand \EOS [0]{\spacefactor3000\relax}%
	\providecommand \BibitemShut  [1]{\csname bibitem#1\endcsname}%
	\let\auto@bib@innerbib\@empty
	\bibitem [{\citenamefont {Krans}\ \emph {et~al.}(1993)\citenamefont {Krans},
	  \citenamefont {Muller}, \citenamefont {Yanson}, \citenamefont {Govaert},
	  \citenamefont {Hesper},\ and\ \citenamefont {van Ruitenbeek}}]{Krans1993}%
	  \BibitemOpen
	  \bibfield  {author} {\bibinfo {author} {\bibfnamefont {J.~M.}\ \bibnamefont
	  {Krans}}, \bibinfo {author} {\bibfnamefont {C.~J.}\ \bibnamefont {Muller}},
	  \bibinfo {author} {\bibfnamefont {I.~K.}\ \bibnamefont {Yanson}}, \bibinfo
	  {author} {\bibfnamefont {T.~C.~M.}\ \bibnamefont {Govaert}}, \bibinfo
	  {author} {\bibfnamefont {R.}~\bibnamefont {Hesper}}, \ and\ \bibinfo {author}
	  {\bibfnamefont {J.~M.}\ \bibnamefont {van Ruitenbeek}},\ }\href {\doibase
	  10.1103/PhysRevB.48.14721} {\bibfield  {journal} {\bibinfo  {journal}
	  {Physical Review B}\ }\textbf {\bibinfo {volume} {48}},\ \bibinfo {pages}
	  {14721} (\bibinfo {year} {1993})}\BibitemShut {NoStop}%
	\bibitem [{\citenamefont {Champagne}\ \emph {et~al.}(2005)\citenamefont
	  {Champagne}, \citenamefont {Pasupathy},\ and\ \citenamefont
	  {Ralph}}]{Champagne2005}%
	  \BibitemOpen
	  \bibfield  {author} {\bibinfo {author} {\bibfnamefont {A.~R.}\ \bibnamefont
	  {Champagne}}, \bibinfo {author} {\bibfnamefont {A.~N.}\ \bibnamefont
	  {Pasupathy}}, \ and\ \bibinfo {author} {\bibfnamefont {D.~C.}\ \bibnamefont
	  {Ralph}},\ }\href {\doibase 10.1021/nl0480619} {\bibfield  {journal}
	  {\bibinfo  {journal} {Nano Letters}\ }\textbf {\bibinfo {volume} {5}},\
	  \bibinfo {pages} {305} (\bibinfo {year} {2005})}\BibitemShut {NoStop}%
	\bibitem [{\citenamefont {Hauptmann}\ \emph {et~al.}(2008)\citenamefont
	  {Hauptmann}, \citenamefont {Paaske},\ and\ \citenamefont
	  {Lindelof}}]{Hauptmann2008}%
	  \BibitemOpen
	  \bibfield  {author} {\bibinfo {author} {\bibfnamefont {J.~R.}\ \bibnamefont
	  {Hauptmann}}, \bibinfo {author} {\bibfnamefont {J.}~\bibnamefont {Paaske}}, \
	  and\ \bibinfo {author} {\bibfnamefont {P.~E.}\ \bibnamefont {Lindelof}},\
	  }\href {\doibase 10.1038/nphys931} {\bibfield  {journal} {\bibinfo  {journal}
	  {Nature Physics}\ }\textbf {\bibinfo {volume} {4}},\ \bibinfo {pages} {373}
	  (\bibinfo {year} {2008})}\BibitemShut {NoStop}%
	\bibitem [{\citenamefont {Egle}\ \emph {et~al.}(2010)\citenamefont {Egle},
	  \citenamefont {Bacca}, \citenamefont {Pernau}, \citenamefont {Huefner},
	  \citenamefont {Hinzke}, \citenamefont {Nowak},\ and\ \citenamefont
	  {Scheer}}]{Egle2010}%
	  \BibitemOpen
	  \bibfield  {author} {\bibinfo {author} {\bibfnamefont {S.}~\bibnamefont
	  {Egle}}, \bibinfo {author} {\bibfnamefont {C.}~\bibnamefont {Bacca}},
	  \bibinfo {author} {\bibfnamefont {H.-F.}\ \bibnamefont {Pernau}}, \bibinfo
	  {author} {\bibfnamefont {M.}~\bibnamefont {Huefner}}, \bibinfo {author}
	  {\bibfnamefont {D.}~\bibnamefont {Hinzke}}, \bibinfo {author} {\bibfnamefont
	  {U.}~\bibnamefont {Nowak}}, \ and\ \bibinfo {author} {\bibfnamefont
	  {E.}~\bibnamefont {Scheer}},\ }\href {\doibase 10.1103/PhysRevB.81.134402}
	  {\bibfield  {journal} {\bibinfo  {journal} {Physical Review B}\ }\textbf
	  {\bibinfo {volume} {81}},\ \bibinfo {pages} {134402} (\bibinfo {year}
	  {2010})}\BibitemShut {NoStop}%
	\bibitem [{\citenamefont {Park}\ \emph {et~al.}(1999)\citenamefont {Park},
	  \citenamefont {Lim}, \citenamefont {Alivisatos}, \citenamefont {Park},\ and\
	  \citenamefont {McEuen}}]{Park1999}%
	  \BibitemOpen
	  \bibfield  {author} {\bibinfo {author} {\bibfnamefont {H.}~\bibnamefont
	  {Park}}, \bibinfo {author} {\bibfnamefont {A.~K.~L.}\ \bibnamefont {Lim}},
	  \bibinfo {author} {\bibfnamefont {a.~P.}\ \bibnamefont {Alivisatos}},
	  \bibinfo {author} {\bibfnamefont {J.}~\bibnamefont {Park}}, \ and\ \bibinfo
	  {author} {\bibfnamefont {P.~L.}\ \bibnamefont {McEuen}},\ }\href {\doibase
	  10.1063/1.124354} {\bibfield  {journal} {\bibinfo  {journal} {Applied Physics
	  Letters}\ }\textbf {\bibinfo {volume} {75}},\ \bibinfo {pages} {301}
	  (\bibinfo {year} {1999})}\BibitemShut {NoStop}%
	\bibitem [{\citenamefont {Brandbyge}\ \emph {et~al.}(1995)\citenamefont
	  {Brandbyge}, \citenamefont {Schi{\o}tz}, \citenamefont {S{\o}rensen},
	  \citenamefont {Stoltze}, \citenamefont {Jacobsen}, \citenamefont
	  {N{\o}rskov}, \citenamefont {Olesen}, \citenamefont {Laegsgaard},
	  \citenamefont {Stensgaard},\ and\ \citenamefont
	  {Besenbacher}}]{Brandbyge1995}%
	  \BibitemOpen
	  \bibfield  {author} {\bibinfo {author} {\bibfnamefont {M.}~\bibnamefont
	  {Brandbyge}}, \bibinfo {author} {\bibfnamefont {J.}~\bibnamefont
	  {Schi{\o}tz}}, \bibinfo {author} {\bibfnamefont {M.~R.}\ \bibnamefont
	  {S{\o}rensen}}, \bibinfo {author} {\bibfnamefont {P.}~\bibnamefont
	  {Stoltze}}, \bibinfo {author} {\bibfnamefont {K.~W.}\ \bibnamefont
	  {Jacobsen}}, \bibinfo {author} {\bibfnamefont {J.~K.}\ \bibnamefont
	  {N{\o}rskov}}, \bibinfo {author} {\bibfnamefont {L.}~\bibnamefont {Olesen}},
	  \bibinfo {author} {\bibfnamefont {E.}~\bibnamefont {Laegsgaard}}, \bibinfo
	  {author} {\bibfnamefont {I.}~\bibnamefont {Stensgaard}}, \ and\ \bibinfo
	  {author} {\bibfnamefont {F.}~\bibnamefont {Besenbacher}},\ }\href {\doibase
	  10.1103/PhysRevB.52.8499} {\bibfield  {journal} {\bibinfo  {journal}
	  {Physical Review B}\ }\textbf {\bibinfo {volume} {52}},\ \bibinfo {pages}
	  {8499} (\bibinfo {year} {1995})}\BibitemShut {NoStop}%
	\bibitem [{\citenamefont {Bernand-Mantel}\ \emph {et~al.}(2006)\citenamefont
	  {Bernand-Mantel}, \citenamefont {Seneor}, \citenamefont {Lidgi},
	  \citenamefont {Mu{\~{n}}oz}, \citenamefont {Cros}, \citenamefont {Fusil},
	  \citenamefont {Bouzehouane}, \citenamefont {Deranlot}, \citenamefont
	  {Vaures}, \citenamefont {Petroff},\ and\ \citenamefont
	  {Fert}}]{Bernand-Mantel2006}%
	  \BibitemOpen
	  \bibfield  {author} {\bibinfo {author} {\bibfnamefont {A.}~\bibnamefont
	  {Bernand-Mantel}}, \bibinfo {author} {\bibfnamefont {P.}~\bibnamefont
	  {Seneor}}, \bibinfo {author} {\bibfnamefont {N.}~\bibnamefont {Lidgi}},
	  \bibinfo {author} {\bibfnamefont {M.}~\bibnamefont {Mu{\~{n}}oz}}, \bibinfo
	  {author} {\bibfnamefont {V.}~\bibnamefont {Cros}}, \bibinfo {author}
	  {\bibfnamefont {S.}~\bibnamefont {Fusil}}, \bibinfo {author} {\bibfnamefont
	  {K.}~\bibnamefont {Bouzehouane}}, \bibinfo {author} {\bibfnamefont
	  {C.}~\bibnamefont {Deranlot}}, \bibinfo {author} {\bibfnamefont
	  {A.}~\bibnamefont {Vaures}}, \bibinfo {author} {\bibfnamefont
	  {F.}~\bibnamefont {Petroff}}, \ and\ \bibinfo {author} {\bibfnamefont
	  {A.}~\bibnamefont {Fert}},\ }\href {\doibase 10.1063/1.2236293} {\bibfield
	  {journal} {\bibinfo  {journal} {Applied Physics Letters}\ }\textbf {\bibinfo
	  {volume} {89}},\ \bibinfo {pages} {062502} (\bibinfo {year}
	  {2006})}\BibitemShut {NoStop}%
	\bibitem [{\citenamefont {Franklin}\ \emph {et~al.}(2012)\citenamefont
	  {Franklin}, \citenamefont {Luisier}, \citenamefont {Han}, \citenamefont
	  {Tulevski}, \citenamefont {Breslin}, \citenamefont {Gignac}, \citenamefont
	  {Lundstrom},\ and\ \citenamefont {Haensch}}]{Franklin2012}%
	  \BibitemOpen
	  \bibfield  {author} {\bibinfo {author} {\bibfnamefont {A.~D.}\ \bibnamefont
	  {Franklin}}, \bibinfo {author} {\bibfnamefont {M.}~\bibnamefont {Luisier}},
	  \bibinfo {author} {\bibfnamefont {S.-J.}\ \bibnamefont {Han}}, \bibinfo
	  {author} {\bibfnamefont {G.}~\bibnamefont {Tulevski}}, \bibinfo {author}
	  {\bibfnamefont {C.~M.}\ \bibnamefont {Breslin}}, \bibinfo {author}
	  {\bibfnamefont {L.}~\bibnamefont {Gignac}}, \bibinfo {author} {\bibfnamefont
	  {M.~S.}\ \bibnamefont {Lundstrom}}, \ and\ \bibinfo {author} {\bibfnamefont
	  {W.}~\bibnamefont {Haensch}},\ }\href {\doibase 10.1021/nl203701g} {\bibfield
	   {journal} {\bibinfo  {journal} {Nano Letters}\ }\textbf {\bibinfo {volume}
	  {12}},\ \bibinfo {pages} {758} (\bibinfo {year} {2012})}\BibitemShut
	  {NoStop}%
	\bibitem [{\citenamefont {Parekh}\ \emph {et~al.}(2012)\citenamefont {Parekh},
	  \citenamefont {Beaumont}, \citenamefont {Beauvais},\ and\ \citenamefont
	  {Drouin}}]{Parekh2012}%
	  \BibitemOpen
	  \bibfield  {author} {\bibinfo {author} {\bibfnamefont {R.}~\bibnamefont
	  {Parekh}}, \bibinfo {author} {\bibfnamefont {A.}~\bibnamefont {Beaumont}},
	  \bibinfo {author} {\bibfnamefont {J.}~\bibnamefont {Beauvais}}, \ and\
	  \bibinfo {author} {\bibfnamefont {D.}~\bibnamefont {Drouin}},\ }\href
	  {\doibase 10.1109/TED.2012.2183374} {\bibfield  {journal} {\bibinfo
	  {journal} {IEEE Transactions on Electron Devices}\ }\textbf {\bibinfo
	  {volume} {59}},\ \bibinfo {pages} {918} (\bibinfo {year} {2012})}\BibitemShut
	  {NoStop}%
	\bibitem [{\citenamefont {Lin}\ \emph {et~al.}(2013)\citenamefont {Lin},
	  \citenamefont {Perez-Barraza}, \citenamefont {Husain}, \citenamefont
	  {Alkhalil}, \citenamefont {Lambert}, \citenamefont {Williams}, \citenamefont
	  {Ferguson}, \citenamefont {Chong},\ and\ \citenamefont {Mizuta}}]{Lin2013}%
	  \BibitemOpen
	  \bibfield  {author} {\bibinfo {author} {\bibfnamefont {Y.~P.}\ \bibnamefont
	  {Lin}}, \bibinfo {author} {\bibfnamefont {J.~I.}\ \bibnamefont
	  {Perez-Barraza}}, \bibinfo {author} {\bibfnamefont {M.~K.}\ \bibnamefont
	  {Husain}}, \bibinfo {author} {\bibfnamefont {F.~M.}\ \bibnamefont
	  {Alkhalil}}, \bibinfo {author} {\bibfnamefont {N.}~\bibnamefont {Lambert}},
	  \bibinfo {author} {\bibfnamefont {D.~A.}\ \bibnamefont {Williams}}, \bibinfo
	  {author} {\bibfnamefont {A.~J.}\ \bibnamefont {Ferguson}}, \bibinfo {author}
	  {\bibfnamefont {H.~M.~H.}\ \bibnamefont {Chong}}, \ and\ \bibinfo {author}
	  {\bibfnamefont {H.}~\bibnamefont {Mizuta}},\ }\href {\doibase
	  10.1109/TNANO.2013.2263964} {\bibfield  {journal} {\bibinfo  {journal} {IEEE
	  Transactions on Nanotechnology}\ }\textbf {\bibinfo {volume} {12}},\ \bibinfo
	  {pages} {897} (\bibinfo {year} {2013})}\BibitemShut {NoStop}%
	\bibitem [{\citenamefont {Pribiag}\ \emph {et~al.}(2013)\citenamefont
	  {Pribiag}, \citenamefont {Nadj-Perge}, \citenamefont {Frolov}, \citenamefont
	  {van~den Berg}, \citenamefont {van Weperen}, \citenamefont {Plissard},
	  \citenamefont {Bakkers},\ and\ \citenamefont {Kouwenhoven}}]{Pribiag2013}%
	  \BibitemOpen
	  \bibfield  {author} {\bibinfo {author} {\bibfnamefont {V.~S.}\ \bibnamefont
	  {Pribiag}}, \bibinfo {author} {\bibfnamefont {S.}~\bibnamefont {Nadj-Perge}},
	  \bibinfo {author} {\bibfnamefont {S.~M.}\ \bibnamefont {Frolov}}, \bibinfo
	  {author} {\bibfnamefont {J.~W.~G.}\ \bibnamefont {van~den Berg}}, \bibinfo
	  {author} {\bibfnamefont {I.}~\bibnamefont {van Weperen}}, \bibinfo {author}
	  {\bibfnamefont {S.~R.}\ \bibnamefont {Plissard}}, \bibinfo {author}
	  {\bibfnamefont {E.~P. A.~M.}\ \bibnamefont {Bakkers}}, \ and\ \bibinfo
	  {author} {\bibfnamefont {L.~P.}\ \bibnamefont {Kouwenhoven}},\ }\href
	  {\doibase 10.1038/nnano.2013.5} {\bibfield  {journal} {\bibinfo  {journal}
	  {Nature Nanotechnology}\ }\textbf {\bibinfo {volume} {8}},\ \bibinfo {pages}
	  {170} (\bibinfo {year} {2013})},\ \Eprint {http://arxiv.org/abs/1302.2648}
	  {arXiv:1302.2648} \BibitemShut {NoStop}%
	\bibitem [{\citenamefont {Desai}\ \emph {et~al.}(2016)\citenamefont {Desai},
	  \citenamefont {Madhvapathy}, \citenamefont {Sachid}, \citenamefont {Llinas},
	  \citenamefont {Wang}, \citenamefont {Ahn}, \citenamefont {Pitner},
	  \citenamefont {Kim}, \citenamefont {Bokor}, \citenamefont {Hu}, \citenamefont
	  {Wong},\ and\ \citenamefont {Javey}}]{Desai2016}%
	  \BibitemOpen
	  \bibfield  {author} {\bibinfo {author} {\bibfnamefont {S.~B.}\ \bibnamefont
	  {Desai}}, \bibinfo {author} {\bibfnamefont {S.~R.}\ \bibnamefont
	  {Madhvapathy}}, \bibinfo {author} {\bibfnamefont {A.~B.}\ \bibnamefont
	  {Sachid}}, \bibinfo {author} {\bibfnamefont {J.~P.}\ \bibnamefont {Llinas}},
	  \bibinfo {author} {\bibfnamefont {Q.}~\bibnamefont {Wang}}, \bibinfo {author}
	  {\bibfnamefont {G.~H.}\ \bibnamefont {Ahn}}, \bibinfo {author} {\bibfnamefont
	  {G.}~\bibnamefont {Pitner}}, \bibinfo {author} {\bibfnamefont {M.~J.}\
	  \bibnamefont {Kim}}, \bibinfo {author} {\bibfnamefont {J.}~\bibnamefont
	  {Bokor}}, \bibinfo {author} {\bibfnamefont {C.}~\bibnamefont {Hu}}, \bibinfo
	  {author} {\bibfnamefont {H.-S.~P.}\ \bibnamefont {Wong}}, \ and\ \bibinfo
	  {author} {\bibfnamefont {A.}~\bibnamefont {Javey}},\ }\href {\doibase
	  10.1126/science.aah4698} {\bibfield  {journal} {\bibinfo  {journal}
	  {Science}\ }\textbf {\bibinfo {volume} {354}},\ \bibinfo {pages} {99}
	  (\bibinfo {year} {2016})}\BibitemShut {NoStop}%
	\bibitem [{\citenamefont {Averin}\ and\ \citenamefont
	  {Likharev}(1986)}]{Averin1986}%
	  \BibitemOpen
	  \bibfield  {author} {\bibinfo {author} {\bibfnamefont {D.~V.}\ \bibnamefont
	  {Averin}}\ and\ \bibinfo {author} {\bibfnamefont {K.~K.}\ \bibnamefont
	  {Likharev}},\ }\href {\doibase 10.1007/BF00683469} {\bibfield  {journal}
	  {\bibinfo  {journal} {Journal of Low Temperature Physics}\ }\textbf {\bibinfo
	  {volume} {62}},\ \bibinfo {pages} {345} (\bibinfo {year} {1986})}\BibitemShut
	  {NoStop}%
	\bibitem [{\citenamefont {Schelp}\ \emph {et~al.}(1997)\citenamefont {Schelp},
	  \citenamefont {Fert}, \citenamefont {Fettar}, \citenamefont {Holody},
	  \citenamefont {Lee}, \citenamefont {Maurice}, \citenamefont {Petroff},\ and\
	  \citenamefont {Vaur{\`{e}}s}}]{Schelp1997}%
	  \BibitemOpen
	  \bibfield  {author} {\bibinfo {author} {\bibfnamefont {L.~F.}\ \bibnamefont
	  {Schelp}}, \bibinfo {author} {\bibfnamefont {A.}~\bibnamefont {Fert}},
	  \bibinfo {author} {\bibfnamefont {F.}~\bibnamefont {Fettar}}, \bibinfo
	  {author} {\bibfnamefont {P.}~\bibnamefont {Holody}}, \bibinfo {author}
	  {\bibfnamefont {S.~F.}\ \bibnamefont {Lee}}, \bibinfo {author} {\bibfnamefont
	  {J.~L.}\ \bibnamefont {Maurice}}, \bibinfo {author} {\bibfnamefont
	  {F.}~\bibnamefont {Petroff}}, \ and\ \bibinfo {author} {\bibfnamefont
	  {A.}~\bibnamefont {Vaur{\`{e}}s}},\ }\href {\doibase
	  10.1103/PhysRevB.56.R5747} {\bibfield  {journal} {\bibinfo  {journal}
	  {Physical Review B}\ }\textbf {\bibinfo {volume} {56}},\ \bibinfo {pages}
	  {R5747} (\bibinfo {year} {1997})}\BibitemShut {NoStop}%
	\bibitem [{\citenamefont {Takahashi}\ and\ \citenamefont
	  {Maekawa}(1998)}]{Takahashi1998}%
	  \BibitemOpen
	  \bibfield  {author} {\bibinfo {author} {\bibfnamefont {S.}~\bibnamefont
	  {Takahashi}}\ and\ \bibinfo {author} {\bibfnamefont {S.}~\bibnamefont
	  {Maekawa}},\ }\href {\doibase 10.1103/PhysRevLett.80.1758} {\bibfield
	  {journal} {\bibinfo  {journal} {Physical Review Letters}\ }\textbf {\bibinfo
	  {volume} {80}},\ \bibinfo {pages} {1758} (\bibinfo {year}
	  {1998})}\BibitemShut {NoStop}%
	\bibitem [{\citenamefont {Likharev}(1999)}]{Likharev1999}%
	  \BibitemOpen
	  \bibfield  {author} {\bibinfo {author} {\bibfnamefont {K.}~\bibnamefont
	  {Likharev}},\ }\href {\doibase 10.1109/5.752518} {\bibfield  {journal}
	  {\bibinfo  {journal} {Proceedings of the IEEE}\ }\textbf {\bibinfo {volume}
	  {87}},\ \bibinfo {pages} {606} (\bibinfo {year} {1999})}\BibitemShut
	  {NoStop}%
	\bibitem [{\citenamefont {Rocha}\ \emph {et~al.}(2007)\citenamefont {Rocha},
	  \citenamefont {Archer},\ and\ \citenamefont {Sanvito}}]{Rocha2007}%
	  \BibitemOpen
	  \bibfield  {author} {\bibinfo {author} {\bibfnamefont {A.~R.}\ \bibnamefont
	  {Rocha}}, \bibinfo {author} {\bibfnamefont {T.}~\bibnamefont {Archer}}, \
	  and\ \bibinfo {author} {\bibfnamefont {S.}~\bibnamefont {Sanvito}},\ }\href
	  {\doibase 10.1103/PhysRevB.76.054435} {\bibfield  {journal} {\bibinfo
	  {journal} {Physical Review B}\ }\textbf {\bibinfo {volume} {76}},\ \bibinfo
	  {pages} {054435} (\bibinfo {year} {2007})}\BibitemShut {NoStop}%
	\bibitem [{\citenamefont {Pontes}\ \emph {et~al.}(2008)\citenamefont {Pontes},
	  \citenamefont {da~Silva}, \citenamefont {Fazzio},\ and\ \citenamefont
	  {da~Silva}}]{Pontes2008}%
	  \BibitemOpen
	  \bibfield  {author} {\bibinfo {author} {\bibfnamefont {R.~B.}\ \bibnamefont
	  {Pontes}}, \bibinfo {author} {\bibfnamefont {E.~Z.}\ \bibnamefont
	  {da~Silva}}, \bibinfo {author} {\bibfnamefont {A.}~\bibnamefont {Fazzio}}, \
	  and\ \bibinfo {author} {\bibfnamefont {A.~J.~R.}\ \bibnamefont {da~Silva}},\
	  }\href {\doibase 10.1021/ja8020457} {\bibfield  {journal} {\bibinfo
	  {journal} {Journal of the American Chemical Society}\ }\textbf {\bibinfo
	  {volume} {130}},\ \bibinfo {pages} {9897} (\bibinfo {year}
	  {2008})}\BibitemShut {NoStop}%
	\bibitem [{\citenamefont {Smogunov}\ \emph {et~al.}(2008)\citenamefont
	  {Smogunov}, \citenamefont {{Dal Corso}}, \citenamefont {Delin}, \citenamefont
	  {Weht},\ and\ \citenamefont {Tosatti}}]{Smogunov2008}%
	  \BibitemOpen
	  \bibfield  {author} {\bibinfo {author} {\bibfnamefont {A.}~\bibnamefont
	  {Smogunov}}, \bibinfo {author} {\bibfnamefont {A.}~\bibnamefont {{Dal
	  Corso}}}, \bibinfo {author} {\bibfnamefont {A.}~\bibnamefont {Delin}},
	  \bibinfo {author} {\bibfnamefont {R.}~\bibnamefont {Weht}}, \ and\ \bibinfo
	  {author} {\bibfnamefont {E.}~\bibnamefont {Tosatti}},\ }\href {\doibase
	  10.1038/nnano.2007.419} {\bibfield  {journal} {\bibinfo  {journal} {Nature
	  Nanotechnology}\ }\textbf {\bibinfo {volume} {3}},\ \bibinfo {pages} {22}
	  (\bibinfo {year} {2008})}\BibitemShut {NoStop}%
	\bibitem [{\citenamefont {Kim}\ \emph {et~al.}(2012)\citenamefont {Kim},
	  \citenamefont {Hashimto}, \citenamefont {Iye},\ and\ \citenamefont
	  {Katsumoto}}]{Kim2012}%
	  \BibitemOpen
	  \bibfield  {author} {\bibinfo {author} {\bibfnamefont {S.~W.}\ \bibnamefont
	  {Kim}}, \bibinfo {author} {\bibfnamefont {Y.}~\bibnamefont {Hashimto}},
	  \bibinfo {author} {\bibfnamefont {Y.}~\bibnamefont {Iye}}, \ and\ \bibinfo
	  {author} {\bibfnamefont {S.}~\bibnamefont {Katsumoto}},\ }\href {\doibase
	  10.1088/1742-6596/400/4/042032} {\bibfield  {journal} {\bibinfo  {journal}
	  {Journal of Physics: Conference Series}\ }\textbf {\bibinfo {volume} {400}},\
	  \bibinfo {pages} {042032} (\bibinfo {year} {2012})}\BibitemShut {NoStop}%
	\bibitem [{\citenamefont {Tsymbal}\ \emph {et~al.}(2003)\citenamefont
	  {Tsymbal}, \citenamefont {Mryasov},\ and\ \citenamefont
	  {LeClair}}]{Tsymbal2003}%
	  \BibitemOpen
	  \bibfield  {author} {\bibinfo {author} {\bibfnamefont {E.~Y.}\ \bibnamefont
	  {Tsymbal}}, \bibinfo {author} {\bibfnamefont {O.~N.}\ \bibnamefont
	  {Mryasov}}, \ and\ \bibinfo {author} {\bibfnamefont {P.~R.}\ \bibnamefont
	  {LeClair}},\ }\href {\doibase 10.1088/0953-8984/15/4/201} {\bibfield
	  {journal} {\bibinfo  {journal} {Journal of Physics: Condensed Matter}\
	  }\textbf {\bibinfo {volume} {15}},\ \bibinfo {pages} {R109} (\bibinfo {year}
	  {2003})}\BibitemShut {NoStop}%
	\bibitem [{\citenamefont {Requist}\ \emph {et~al.}(2016)\citenamefont
	  {Requist}, \citenamefont {Baruselli}, \citenamefont {Smogunov}, \citenamefont
	  {Fabrizio}, \citenamefont {Modesti},\ and\ \citenamefont
	  {Tosatti}}]{Requist2016}%
	  \BibitemOpen
	  \bibfield  {author} {\bibinfo {author} {\bibfnamefont {R.}~\bibnamefont
	  {Requist}}, \bibinfo {author} {\bibfnamefont {P.~P.}\ \bibnamefont
	  {Baruselli}}, \bibinfo {author} {\bibfnamefont {A.}~\bibnamefont {Smogunov}},
	  \bibinfo {author} {\bibfnamefont {M.}~\bibnamefont {Fabrizio}}, \bibinfo
	  {author} {\bibfnamefont {S.}~\bibnamefont {Modesti}}, \ and\ \bibinfo
	  {author} {\bibfnamefont {E.}~\bibnamefont {Tosatti}},\ }\href {\doibase
	  10.1038/nnano.2016.55} {\bibfield  {journal} {\bibinfo  {journal} {Nature
	  Nanotechnology}\ }\textbf {\bibinfo {volume} {11}},\ \bibinfo {pages} {499}
	  (\bibinfo {year} {2016})}\BibitemShut {NoStop}%
	\bibitem [{\citenamefont {Tao}\ \emph {et~al.}(2009)\citenamefont {Tao},
	  \citenamefont {Stepanyuk}, \citenamefont {Hergert}, \citenamefont {Rungger},
	  \citenamefont {Sanvito},\ and\ \citenamefont {Bruno}}]{Tao2009}%
	  \BibitemOpen
	  \bibfield  {author} {\bibinfo {author} {\bibfnamefont {K.}~\bibnamefont
	  {Tao}}, \bibinfo {author} {\bibfnamefont {V.~S.}\ \bibnamefont {Stepanyuk}},
	  \bibinfo {author} {\bibfnamefont {W.}~\bibnamefont {Hergert}}, \bibinfo
	  {author} {\bibfnamefont {I.}~\bibnamefont {Rungger}}, \bibinfo {author}
	  {\bibfnamefont {S.}~\bibnamefont {Sanvito}}, \ and\ \bibinfo {author}
	  {\bibfnamefont {P.}~\bibnamefont {Bruno}},\ }\href {\doibase
	  10.1103/PhysRevLett.103.057202} {\bibfield  {journal} {\bibinfo  {journal}
	  {Phys. Rev. Lett.}\ }\textbf {\bibinfo {volume} {103}},\ \bibinfo {pages}
	  {057202} (\bibinfo {year} {2009})}\BibitemShut {NoStop}%
	\bibitem [{\citenamefont {Tsukamoto}\ \emph {et~al.}(2013)\citenamefont
	  {Tsukamoto}, \citenamefont {Caciuc}, \citenamefont {Atodiresei},\ and\
	  \citenamefont {Bl{\"{u}}gel}}]{Tsukamoto2013}%
	  \BibitemOpen
	  \bibfield  {author} {\bibinfo {author} {\bibfnamefont {S.}~\bibnamefont
	  {Tsukamoto}}, \bibinfo {author} {\bibfnamefont {V.}~\bibnamefont {Caciuc}},
	  \bibinfo {author} {\bibfnamefont {N.}~\bibnamefont {Atodiresei}}, \ and\
	  \bibinfo {author} {\bibfnamefont {S.}~\bibnamefont {Bl{\"{u}}gel}},\ }\href
	  {\doibase 10.1103/PhysRevB.88.125436} {\bibfield  {journal} {\bibinfo
	  {journal} {Physical Review B}\ }\textbf {\bibinfo {volume} {88}},\ \bibinfo
	  {pages} {125436} (\bibinfo {year} {2013})}\BibitemShut {NoStop}%
	\bibitem [{\citenamefont {Sivkov}\ \emph
	  {et~al.}(2014{\natexlab{a}})\citenamefont {Sivkov}, \citenamefont {Brovko},\
	  and\ \citenamefont {Stepanyuk}}]{Sivkov2014a}%
	  \BibitemOpen
	  \bibfield  {author} {\bibinfo {author} {\bibfnamefont {I.~N.}\ \bibnamefont
	  {Sivkov}}, \bibinfo {author} {\bibfnamefont {O.~O.}\ \bibnamefont {Brovko}},
	  \ and\ \bibinfo {author} {\bibfnamefont {V.~S.}\ \bibnamefont {Stepanyuk}},\
	  }\href {\doibase 10.1103/PhysRevB.89.195419} {\bibfield  {journal} {\bibinfo
	  {journal} {Physical Review B}\ }\textbf {\bibinfo {volume} {89}},\ \bibinfo
	  {pages} {195419} (\bibinfo {year} {2014}{\natexlab{a}})}\BibitemShut
	  {NoStop}%
	\bibitem [{\citenamefont {Obermair}\ \emph {et~al.}(2010)\citenamefont
	  {Obermair}, \citenamefont {Xie},\ and\ \citenamefont
	  {Schimmel}}]{Obermair2010}%
	  \BibitemOpen
	  \bibfield  {author} {\bibinfo {author} {\bibfnamefont {C.}~\bibnamefont
	  {Obermair}}, \bibinfo {author} {\bibfnamefont {F.-Q.}\ \bibnamefont {Xie}}, \
	  and\ \bibinfo {author} {\bibfnamefont {T.}~\bibnamefont {Schimmel}},\ }\href
	  {\doibase 10.1051/epn/2010403} {\bibfield  {journal} {\bibinfo  {journal}
	  {Europhysics News}\ }\textbf {\bibinfo {volume} {41}},\ \bibinfo {pages} {25}
	  (\bibinfo {year} {2010})}\BibitemShut {NoStop}%
	\bibitem [{\citenamefont {Osorio}\ \emph {et~al.}(2008)\citenamefont {Osorio},
	  \citenamefont {Bj{\o}rnholm}, \citenamefont {Lehn}, \citenamefont {Ruben},\
	  and\ \citenamefont {van~der Zant}}]{Osorio2008}%
	  \BibitemOpen
	  \bibfield  {author} {\bibinfo {author} {\bibfnamefont {E.~A.}\ \bibnamefont
	  {Osorio}}, \bibinfo {author} {\bibfnamefont {T.}~\bibnamefont
	  {Bj{\o}rnholm}}, \bibinfo {author} {\bibfnamefont {J.-M.}\ \bibnamefont
	  {Lehn}}, \bibinfo {author} {\bibfnamefont {M.}~\bibnamefont {Ruben}}, \ and\
	  \bibinfo {author} {\bibfnamefont {H.~S.~J.}\ \bibnamefont {van~der Zant}},\
	  }\href {\doibase 10.1088/0953-8984/20/37/374121} {\bibfield  {journal}
	  {\bibinfo  {journal} {Journal of Physics: Condensed Matter}\ }\textbf
	  {\bibinfo {volume} {20}},\ \bibinfo {pages} {374121} (\bibinfo {year}
	  {2008})}\BibitemShut {NoStop}%
	\bibitem [{\citenamefont {Lee}\ \emph {et~al.}(2013)\citenamefont {Lee},
	  \citenamefont {Park}, \citenamefont {Song},\ and\ \citenamefont
	  {Kim}}]{Lee2013}%
	  \BibitemOpen
	  \bibfield  {author} {\bibinfo {author} {\bibfnamefont {J.~S.}\ \bibnamefont
	  {Lee}}, \bibinfo {author} {\bibfnamefont {J.-W.}\ \bibnamefont {Park}},
	  \bibinfo {author} {\bibfnamefont {J.~Y.}\ \bibnamefont {Song}}, \ and\
	  \bibinfo {author} {\bibfnamefont {J.}~\bibnamefont {Kim}},\ }\href {\doibase
	  10.1088/0957-4484/24/19/195201} {\bibfield  {journal} {\bibinfo  {journal}
	  {Nanotechnology}\ }\textbf {\bibinfo {volume} {24}},\ \bibinfo {pages}
	  {195201} (\bibinfo {year} {2013})}\BibitemShut {NoStop}%
	\bibitem [{\citenamefont {Modepalli}\ \emph {et~al.}(2016)\citenamefont
	  {Modepalli}, \citenamefont {Jin}, \citenamefont {Park}, \citenamefont {Jo},
	  \citenamefont {Kim}, \citenamefont {Baik}, \citenamefont {Seo}, \citenamefont
	  {Kim},\ and\ \citenamefont {Yoo}}]{Modepalli2016}%
	  \BibitemOpen
	  \bibfield  {author} {\bibinfo {author} {\bibfnamefont {V.}~\bibnamefont
	  {Modepalli}}, \bibinfo {author} {\bibfnamefont {M.-J.}\ \bibnamefont {Jin}},
	  \bibinfo {author} {\bibfnamefont {J.}~\bibnamefont {Park}}, \bibinfo {author}
	  {\bibfnamefont {J.}~\bibnamefont {Jo}}, \bibinfo {author} {\bibfnamefont
	  {J.-H.}\ \bibnamefont {Kim}}, \bibinfo {author} {\bibfnamefont {J.~M.}\
	  \bibnamefont {Baik}}, \bibinfo {author} {\bibfnamefont {C.}~\bibnamefont
	  {Seo}}, \bibinfo {author} {\bibfnamefont {J.}~\bibnamefont {Kim}}, \ and\
	  \bibinfo {author} {\bibfnamefont {J.-W.}\ \bibnamefont {Yoo}},\ }\href
	  {\doibase 10.1021/acsnano.6b00921} {\bibfield  {journal} {\bibinfo  {journal}
	  {ACS Nano}\ }\textbf {\bibinfo {volume} {10}},\ \bibinfo {pages} {4618}
	  (\bibinfo {year} {2016})}\BibitemShut {NoStop}%
	\bibitem [{\citenamefont {Wang}\ \emph {et~al.}(2016)\citenamefont {Wang},
	  \citenamefont {Li}, \citenamefont {Yu}, \citenamefont {Wei}, \citenamefont
	  {Wang},\ and\ \citenamefont {Guo}}]{Wang2016}%
	  \BibitemOpen
	  \bibfield  {author} {\bibinfo {author} {\bibfnamefont {B.}~\bibnamefont
	  {Wang}}, \bibinfo {author} {\bibfnamefont {J.}~\bibnamefont {Li}}, \bibinfo
	  {author} {\bibfnamefont {Y.}~\bibnamefont {Yu}}, \bibinfo {author}
	  {\bibfnamefont {Y.}~\bibnamefont {Wei}}, \bibinfo {author} {\bibfnamefont
	  {J.}~\bibnamefont {Wang}}, \ and\ \bibinfo {author} {\bibfnamefont
	  {H.}~\bibnamefont {Guo}},\ }\href {\doibase 10.1039/C5NR06585B} {\bibfield
	  {journal} {\bibinfo  {journal} {Nanoscale}\ }\textbf {\bibinfo {volume}
	  {8}},\ \bibinfo {pages} {3432} (\bibinfo {year} {2016})}\BibitemShut
	  {NoStop}%
	\bibitem [{\citenamefont {Dasa}\ \emph {et~al.}(2012)\citenamefont {Dasa},
	  \citenamefont {Ignatiev},\ and\ \citenamefont {Stepanyuk}}]{Dasa2012}%
	  \BibitemOpen
	  \bibfield  {author} {\bibinfo {author} {\bibfnamefont {T.~R.}\ \bibnamefont
	  {Dasa}}, \bibinfo {author} {\bibfnamefont {P.~A.}\ \bibnamefont {Ignatiev}},
	  \ and\ \bibinfo {author} {\bibfnamefont {V.~S.}\ \bibnamefont {Stepanyuk}},\
	  }\href {\doibase 10.1103/PhysRevB.85.205447} {\bibfield  {journal} {\bibinfo
	  {journal} {Physical Peview B}\ }\textbf {\bibinfo {volume} {85}},\ \bibinfo
	  {pages} {205447} (\bibinfo {year} {2012})}\BibitemShut {NoStop}%
	\bibitem [{\citenamefont {Smogunov}\ and\ \citenamefont
	  {Dappe}(2015)}]{Smogunov2015}%
	  \BibitemOpen
	  \bibfield  {author} {\bibinfo {author} {\bibfnamefont {A.}~\bibnamefont
	  {Smogunov}}\ and\ \bibinfo {author} {\bibfnamefont {Y.~J.}\ \bibnamefont
	  {Dappe}},\ }\href {\doibase 10.1021/acs.nanolett.5b01004} {\bibfield
	  {journal} {\bibinfo  {journal} {Nano Letters}\ }\textbf {\bibinfo {volume}
	  {15}},\ \bibinfo {pages} {3552} (\bibinfo {year} {2015})}\BibitemShut
	  {NoStop}%
	\bibitem [{\citenamefont {Li}\ \emph {et~al.}(2016)\citenamefont {Li},
	  \citenamefont {Wang}, \citenamefont {Xu}, \citenamefont {Wei},\ and\
	  \citenamefont {Wang}}]{Li2016}%
	  \BibitemOpen
	  \bibfield  {author} {\bibinfo {author} {\bibfnamefont {J.}~\bibnamefont
	  {Li}}, \bibinfo {author} {\bibfnamefont {B.}~\bibnamefont {Wang}}, \bibinfo
	  {author} {\bibfnamefont {F.}~\bibnamefont {Xu}}, \bibinfo {author}
	  {\bibfnamefont {Y.}~\bibnamefont {Wei}}, \ and\ \bibinfo {author}
	  {\bibfnamefont {J.}~\bibnamefont {Wang}},\ }\href {\doibase
	  10.1103/PhysRevB.93.195426} {\bibfield  {journal} {\bibinfo  {journal}
	  {Physical Review B}\ }\textbf {\bibinfo {volume} {93}},\ \bibinfo {pages}
	  {195426} (\bibinfo {year} {2016})}\BibitemShut {NoStop}%
	\bibitem [{\citenamefont {Brovko}\ \emph {et~al.}(2014)\citenamefont {Brovko},
	  \citenamefont {Ruiz-D{\'{i}}az}, \citenamefont {Dasa},\ and\ \citenamefont
	  {Stepanyuk}}]{Brovko2014}%
	  \BibitemOpen
	  \bibfield  {author} {\bibinfo {author} {\bibfnamefont {O.~O.}\ \bibnamefont
	  {Brovko}}, \bibinfo {author} {\bibfnamefont {P.}~\bibnamefont
	  {Ruiz-D{\'{i}}az}}, \bibinfo {author} {\bibfnamefont {T.~R.}\ \bibnamefont
	  {Dasa}}, \ and\ \bibinfo {author} {\bibfnamefont {V.~S.}\ \bibnamefont
	  {Stepanyuk}},\ }\href {\doibase 10.1088/0953-8984/26/9/093001} {\bibfield
	  {journal} {\bibinfo  {journal} {Journal of Physics: Condensed Matter}\
	  }\textbf {\bibinfo {volume} {26}},\ \bibinfo {pages} {093001} (\bibinfo
	  {year} {2014})}\BibitemShut {NoStop}%
	\bibitem [{\citenamefont {Rocha}\ \emph {et~al.}(2005)\citenamefont {Rocha},
	  \citenamefont {Garc{\'{i}}a-su{\'{a}}rez}, \citenamefont {Bailey},
	  \citenamefont {Lambert}, \citenamefont {Ferrer},\ and\ \citenamefont
	  {Sanvito}}]{Rocha2005}%
	  \BibitemOpen
	  \bibfield  {author} {\bibinfo {author} {\bibfnamefont {A.~R.}\ \bibnamefont
	  {Rocha}}, \bibinfo {author} {\bibfnamefont {V.~M.}\ \bibnamefont
	  {Garc{\'{i}}a-su{\'{a}}rez}}, \bibinfo {author} {\bibfnamefont {S.~W.}\
	  \bibnamefont {Bailey}}, \bibinfo {author} {\bibfnamefont {C.~J.}\
	  \bibnamefont {Lambert}}, \bibinfo {author} {\bibfnamefont {J.}~\bibnamefont
	  {Ferrer}}, \ and\ \bibinfo {author} {\bibfnamefont {S.}~\bibnamefont
	  {Sanvito}},\ }\href {\doibase 10.1038/nmat1349} {\bibfield  {journal}
	  {\bibinfo  {journal} {Nature Materials}\ }\textbf {\bibinfo {volume} {4}},\
	  \bibinfo {pages} {335} (\bibinfo {year} {2005})}\BibitemShut {NoStop}%
	\bibitem [{\citenamefont {Rocha}\ \emph {et~al.}(2006)\citenamefont {Rocha},
	  \citenamefont {Garc{\'{i}}a-Su{\'{a}}rez}, \citenamefont {Bailey},
	  \citenamefont {Lambert}, \citenamefont {Ferrer},\ and\ \citenamefont
	  {Sanvito}}]{Rocha2006}%
	  \BibitemOpen
	  \bibfield  {author} {\bibinfo {author} {\bibfnamefont {A.~R.}\ \bibnamefont
	  {Rocha}}, \bibinfo {author} {\bibfnamefont {V.~M.}\ \bibnamefont
	  {Garc{\'{i}}a-Su{\'{a}}rez}}, \bibinfo {author} {\bibfnamefont
	  {S.}~\bibnamefont {Bailey}}, \bibinfo {author} {\bibfnamefont
	  {C.}~\bibnamefont {Lambert}}, \bibinfo {author} {\bibfnamefont
	  {J.}~\bibnamefont {Ferrer}}, \ and\ \bibinfo {author} {\bibfnamefont
	  {S.}~\bibnamefont {Sanvito}},\ }\href {\doibase 10.1103/PhysRevB.73.085414}
	  {\bibfield  {journal} {\bibinfo  {journal} {Physical Review B}\ }\textbf
	  {\bibinfo {volume} {73}},\ \bibinfo {pages} {085414} (\bibinfo {year}
	  {2006})}\BibitemShut {NoStop}%
	\bibitem [{\citenamefont {Rungger}\ and\ \citenamefont
	  {Sanvito}(2008)}]{Rungger2008}%
	  \BibitemOpen
	  \bibfield  {author} {\bibinfo {author} {\bibfnamefont {I.}~\bibnamefont
	  {Rungger}}\ and\ \bibinfo {author} {\bibfnamefont {S.}~\bibnamefont
	  {Sanvito}},\ }\href {\doibase 10.1103/PhysRevB.78.035407} {\bibfield
	  {journal} {\bibinfo  {journal} {Physical Review B}\ }\textbf {\bibinfo
	  {volume} {78}},\ \bibinfo {pages} {035407} (\bibinfo {year}
	  {2008})}\BibitemShut {NoStop}%
	\bibitem [{\citenamefont {Ordej{\'{o}}n}\ \emph {et~al.}(1996)\citenamefont
	  {Ordej{\'{o}}n}, \citenamefont {Artacho},\ and\ \citenamefont
	  {Soler}}]{Ordejon1996}%
	  \BibitemOpen
	  \bibfield  {author} {\bibinfo {author} {\bibfnamefont {P.}~\bibnamefont
	  {Ordej{\'{o}}n}}, \bibinfo {author} {\bibfnamefont {E.}~\bibnamefont
	  {Artacho}}, \ and\ \bibinfo {author} {\bibfnamefont {J.~M.}\ \bibnamefont
	  {Soler}},\ }\href {\doibase 10.1103/PhysRevB.53.R10441} {\bibfield  {journal}
	  {\bibinfo  {journal} {Physical Review B}\ }\textbf {\bibinfo {volume} {53}},\
	  \bibinfo {pages} {R10441} (\bibinfo {year} {1996})}\BibitemShut {NoStop}%
	\bibitem [{\citenamefont {Soler}\ \emph {et~al.}(2002)\citenamefont {Soler},
	  \citenamefont {Artacho}, \citenamefont {Gale}, \citenamefont {Garc{\'{i}}a},
	  \citenamefont {Junquera}, \citenamefont {Ordej{\'{o}}n},\ and\ \citenamefont
	  {S{\'{a}}nchez-Portal}}]{Soler2002}%
	  \BibitemOpen
	  \bibfield  {author} {\bibinfo {author} {\bibfnamefont {J.~M.}\ \bibnamefont
	  {Soler}}, \bibinfo {author} {\bibfnamefont {E.}~\bibnamefont {Artacho}},
	  \bibinfo {author} {\bibfnamefont {J.~D.}\ \bibnamefont {Gale}}, \bibinfo
	  {author} {\bibfnamefont {A.}~\bibnamefont {Garc{\'{i}}a}}, \bibinfo {author}
	  {\bibfnamefont {J.}~\bibnamefont {Junquera}}, \bibinfo {author}
	  {\bibfnamefont {P.}~\bibnamefont {Ordej{\'{o}}n}}, \ and\ \bibinfo {author}
	  {\bibfnamefont {D.}~\bibnamefont {S{\'{a}}nchez-Portal}},\ }\href {\doibase
	  10.1088/0953-8984/14/11/302} {\bibfield  {journal} {\bibinfo  {journal}
	  {Journal of Physics: Condensed Matter}\ }\textbf {\bibinfo {volume} {14}},\
	  \bibinfo {pages} {2745} (\bibinfo {year} {2002})}\BibitemShut {NoStop}%
	\bibitem [{\citenamefont {Monkhorst}\ and\ \citenamefont
	  {Pack}(1976)}]{Monkhorst1976}%
	  \BibitemOpen
	  \bibfield  {author} {\bibinfo {author} {\bibfnamefont {H.~J.}\ \bibnamefont
	  {Monkhorst}}\ and\ \bibinfo {author} {\bibfnamefont {J.~D.}\ \bibnamefont
	  {Pack}},\ }\href {\doibase 10.1103/PhysRevB.13.5188} {\bibfield  {journal}
	  {\bibinfo  {journal} {Phys. Rev. B}\ }\textbf {\bibinfo {volume} {13}},\
	  \bibinfo {pages} {5188} (\bibinfo {year} {1976})}\BibitemShut {NoStop}%
	\bibitem [{\citenamefont {Weisheit}\ \emph {et~al.}(2007)\citenamefont
	  {Weisheit}, \citenamefont {F{\"{a}}hler}, \citenamefont {Marty},
	  \citenamefont {Souche}, \citenamefont {Poinsignon},\ and\ \citenamefont
	  {Givord}}]{Weisheit2007}%
	  \BibitemOpen
	  \bibfield  {author} {\bibinfo {author} {\bibfnamefont {M.}~\bibnamefont
	  {Weisheit}}, \bibinfo {author} {\bibfnamefont {S.}~\bibnamefont
	  {F{\"{a}}hler}}, \bibinfo {author} {\bibfnamefont {A.}~\bibnamefont {Marty}},
	  \bibinfo {author} {\bibfnamefont {Y.}~\bibnamefont {Souche}}, \bibinfo
	  {author} {\bibfnamefont {C.}~\bibnamefont {Poinsignon}}, \ and\ \bibinfo
	  {author} {\bibfnamefont {D.}~\bibnamefont {Givord}},\ }\href {\doibase
	  10.1126/science.1136629} {\bibfield  {journal} {\bibinfo  {journal}
	  {Science}\ }\textbf {\bibinfo {volume} {315}},\ \bibinfo {pages} {349}
	  (\bibinfo {year} {2007})}\BibitemShut {NoStop}%
	\bibitem [{Note1()}]{Note1}%
	  \BibitemOpen
	  \bibinfo {note} {The validity of this approximation is corroborated by the
	  similarity in equilibrium and non-equilibrium conductance curves in Fig.~\ref
	  {fig:iv}}\BibitemShut {NoStop}%
	\bibitem [{\citenamefont {Stepanyuk}\ \emph {et~al.}(2004)\citenamefont
	  {Stepanyuk}, \citenamefont {Bruno}, \citenamefont {Klavsyuk}, \citenamefont
	  {Baranov}, \citenamefont {Hergert}, \citenamefont {Saletsky},\ and\
	  \citenamefont {Mertig}}]{Stepanyuk2004a}%
	  \BibitemOpen
	  \bibfield  {author} {\bibinfo {author} {\bibfnamefont {V.~S.}\ \bibnamefont
	  {Stepanyuk}}, \bibinfo {author} {\bibfnamefont {P.}~\bibnamefont {Bruno}},
	  \bibinfo {author} {\bibfnamefont {A.~L.}\ \bibnamefont {Klavsyuk}}, \bibinfo
	  {author} {\bibfnamefont {A.~N.}\ \bibnamefont {Baranov}}, \bibinfo {author}
	  {\bibfnamefont {W.}~\bibnamefont {Hergert}}, \bibinfo {author} {\bibfnamefont
	  {A.~M.}\ \bibnamefont {Saletsky}}, \ and\ \bibinfo {author} {\bibfnamefont
	  {I.}~\bibnamefont {Mertig}},\ }\href {\doibase 10.1103/PhysRevB.69.033302}
	  {\bibfield  {journal} {\bibinfo  {journal} {Physical Review B}\ }\textbf
	  {\bibinfo {volume} {69}},\ \bibinfo {pages} {033302} (\bibinfo {year}
	  {2004})}\BibitemShut {NoStop}%
	\bibitem [{\citenamefont {Emberly}\ and\ \citenamefont
	  {Kirczenow}(1999)}]{Emberly1999}%
	  \BibitemOpen
	  \bibfield  {author} {\bibinfo {author} {\bibfnamefont {E.~G.}\ \bibnamefont
	  {Emberly}}\ and\ \bibinfo {author} {\bibfnamefont {G.}~\bibnamefont
	  {Kirczenow}},\ }\href {\doibase 10.1103/PhysRevB.60.6028} {\bibfield
	  {journal} {\bibinfo  {journal} {Physical Review B}\ }\textbf {\bibinfo
	  {volume} {60}},\ \bibinfo {pages} {6028} (\bibinfo {year}
	  {1999})}\BibitemShut {NoStop}%
	\bibitem [{\citenamefont {Sivkov}\ \emph
	  {et~al.}(2014{\natexlab{b}})\citenamefont {Sivkov}, \citenamefont {Brovko},
	  \citenamefont {Bazhanov},\ and\ \citenamefont {Stepanyuk}}]{Sivkov2014}%
	  \BibitemOpen
	  \bibfield  {author} {\bibinfo {author} {\bibfnamefont {I.~N.}\ \bibnamefont
	  {Sivkov}}, \bibinfo {author} {\bibfnamefont {O.~O.}\ \bibnamefont {Brovko}},
	  \bibinfo {author} {\bibfnamefont {D.~I.}\ \bibnamefont {Bazhanov}}, \ and\
	  \bibinfo {author} {\bibfnamefont {V.~S.}\ \bibnamefont {Stepanyuk}},\ }\href
	  {\doibase 10.1103/PhysRevB.89.075436} {\bibfield  {journal} {\bibinfo
	  {journal} {Physical Review B}\ }\textbf {\bibinfo {volume} {89}},\ \bibinfo
	  {pages} {075436} (\bibinfo {year} {2014}{\natexlab{b}})}\BibitemShut
	  {NoStop}%
	\bibitem [{\citenamefont {Rau}\ \emph {et~al.}(2014)\citenamefont {Rau},
	  \citenamefont {Baumann}, \citenamefont {Rusponi}, \citenamefont {Donati},
	  \citenamefont {Stepanow}, \citenamefont {Gragnaniello}, \citenamefont
	  {Dreiser}, \citenamefont {Piamonteze}, \citenamefont {Nolting}, \citenamefont
	  {Gangopadhyay}, \citenamefont {Albertini}, \citenamefont {Macfarlane},
	  \citenamefont {Lutz}, \citenamefont {Jones}, \citenamefont {Gambardella},
	  \citenamefont {Heinrich},\ and\ \citenamefont {Brune}}]{Rau2014}%
	  \BibitemOpen
	  \bibfield  {author} {\bibinfo {author} {\bibfnamefont {I.~G.}\ \bibnamefont
	  {Rau}}, \bibinfo {author} {\bibfnamefont {S.}~\bibnamefont {Baumann}},
	  \bibinfo {author} {\bibfnamefont {S.}~\bibnamefont {Rusponi}}, \bibinfo
	  {author} {\bibfnamefont {F.}~\bibnamefont {Donati}}, \bibinfo {author}
	  {\bibfnamefont {S.}~\bibnamefont {Stepanow}}, \bibinfo {author}
	  {\bibfnamefont {L.}~\bibnamefont {Gragnaniello}}, \bibinfo {author}
	  {\bibfnamefont {J.}~\bibnamefont {Dreiser}}, \bibinfo {author} {\bibfnamefont
	  {C.}~\bibnamefont {Piamonteze}}, \bibinfo {author} {\bibfnamefont
	  {F.}~\bibnamefont {Nolting}}, \bibinfo {author} {\bibfnamefont
	  {S.}~\bibnamefont {Gangopadhyay}}, \bibinfo {author} {\bibfnamefont {O.~R.}\
	  \bibnamefont {Albertini}}, \bibinfo {author} {\bibfnamefont {R.~M.}\
	  \bibnamefont {Macfarlane}}, \bibinfo {author} {\bibfnamefont {C.~P.}\
	  \bibnamefont {Lutz}}, \bibinfo {author} {\bibfnamefont {B.~A.}\ \bibnamefont
	  {Jones}}, \bibinfo {author} {\bibfnamefont {P.}~\bibnamefont {Gambardella}},
	  \bibinfo {author} {\bibfnamefont {A.~J.}\ \bibnamefont {Heinrich}}, \ and\
	  \bibinfo {author} {\bibfnamefont {H.}~\bibnamefont {Brune}},\ }\href
	  {\doibase 10.1126/science.1252841} {\bibfield  {journal} {\bibinfo  {journal}
	  {Science}\ }\textbf {\bibinfo {volume} {344}},\ \bibinfo {pages} {988}
	  (\bibinfo {year} {2014})}\BibitemShut {NoStop}%
	\bibitem [{\citenamefont {Tsysar}\ \emph {et~al.}(2012)\citenamefont {Tsysar},
	  \citenamefont {Bazhanov}, \citenamefont {Smelova},\ and\ \citenamefont
	  {Saletsky}}]{Tsysar2012}%
	  \BibitemOpen
	  \bibfield  {author} {\bibinfo {author} {\bibfnamefont {K.~M.}\ \bibnamefont
	  {Tsysar}}, \bibinfo {author} {\bibfnamefont {D.~I.}\ \bibnamefont
	  {Bazhanov}}, \bibinfo {author} {\bibfnamefont {E.~M.}\ \bibnamefont
	  {Smelova}}, \ and\ \bibinfo {author} {\bibfnamefont {A.~M.}\ \bibnamefont
	  {Saletsky}},\ }\href {\doibase 10.1063/1.4738767} {\bibfield  {journal}
	  {\bibinfo  {journal} {Applied Physics Letters}\ }\textbf {\bibinfo {volume}
	  {101}},\ \bibinfo {pages} {1} (\bibinfo {year} {2012})}\BibitemShut {NoStop}%
	\bibitem [{\citenamefont {Lipi{\'{n}}ski}\ and\ \citenamefont
	  {Krychowski}(2010)}]{Lipinski2010}%
	  \BibitemOpen
	  \bibfield  {author} {\bibinfo {author} {\bibfnamefont {S.}~\bibnamefont
	  {Lipi{\'{n}}ski}}\ and\ \bibinfo {author} {\bibfnamefont {D.}~\bibnamefont
	  {Krychowski}},\ }\href {\doibase 10.1103/PhysRevB.81.115327} {\bibfield
	  {journal} {\bibinfo  {journal} {Physical Review B}\ }\textbf {\bibinfo
	  {volume} {81}},\ \bibinfo {pages} {115327} (\bibinfo {year}
	  {2010})}\BibitemShut {NoStop}%
	\bibitem [{\citenamefont {Meyer}\ \emph {et~al.}(2001)\citenamefont {Meyer},
	  \citenamefont {Bartels},\ and\ \citenamefont {Rieder}}]{Meyer2001}%
	  \BibitemOpen
	  \bibfield  {author} {\bibinfo {author} {\bibfnamefont {G.}~\bibnamefont
	  {Meyer}}, \bibinfo {author} {\bibfnamefont {L.}~\bibnamefont {Bartels}}, \
	  and\ \bibinfo {author} {\bibfnamefont {K.~H.}\ \bibnamefont {Rieder}},\
	  }\href {\doibase 10.1016/S0927-0256(00)00205-6} {\bibfield  {journal}
	  {\bibinfo  {journal} {Computational Materials Science}\ }\textbf {\bibinfo
	  {volume} {20}},\ \bibinfo {pages} {443} (\bibinfo {year} {2001})}\BibitemShut
	  {NoStop}%
	\bibitem [{\citenamefont {Khajetoorians}\ \emph {et~al.}(2011)\citenamefont
	  {Khajetoorians}, \citenamefont {Wiebe}, \citenamefont {Chilian},\ and\
	  \citenamefont {Wiesendanger}}]{Khajetoorians2011}%
	  \BibitemOpen
	  \bibfield  {author} {\bibinfo {author} {\bibfnamefont {A.~A.}\ \bibnamefont
	  {Khajetoorians}}, \bibinfo {author} {\bibfnamefont {J.}~\bibnamefont
	  {Wiebe}}, \bibinfo {author} {\bibfnamefont {B.}~\bibnamefont {Chilian}}, \
	  and\ \bibinfo {author} {\bibfnamefont {R.}~\bibnamefont {Wiesendanger}},\
	  }\href {\doibase 10.1126/science.1201725} {\bibfield  {journal} {\bibinfo
	  {journal} {Science}\ }\textbf {\bibinfo {volume} {332}},\ \bibinfo {pages}
	  {1062} (\bibinfo {year} {2011})}\BibitemShut {NoStop}%
	\bibitem [{\citenamefont {Hirjibehedin}\ \emph {et~al.}(2006)\citenamefont
	  {Hirjibehedin}, \citenamefont {Lutz},\ and\ \citenamefont
	  {Heinrich}}]{Hirjibehedin2006}%
	  \BibitemOpen
	  \bibfield  {author} {\bibinfo {author} {\bibfnamefont {C.~F.}\ \bibnamefont
	  {Hirjibehedin}}, \bibinfo {author} {\bibfnamefont {C.~P.}\ \bibnamefont
	  {Lutz}}, \ and\ \bibinfo {author} {\bibfnamefont {A.~J.}\ \bibnamefont
	  {Heinrich}},\ }\href {\doibase 10.1126/science.1125398} {\bibfield  {journal}
	  {\bibinfo  {journal} {Science}\ }\textbf {\bibinfo {volume} {312}},\ \bibinfo
	  {pages} {1021} (\bibinfo {year} {2006})}\BibitemShut {NoStop}%
	\bibitem [{\citenamefont {Otte}\ \emph {et~al.}(2009)\citenamefont {Otte},
	  \citenamefont {Ternes}, \citenamefont {Loth}, \citenamefont {Lutz},
	  \citenamefont {Hirjibehedin},\ and\ \citenamefont {Heinrich}}]{Otte2009}%
	  \BibitemOpen
	  \bibfield  {author} {\bibinfo {author} {\bibfnamefont {A.~F.}\ \bibnamefont
	  {Otte}}, \bibinfo {author} {\bibfnamefont {M.}~\bibnamefont {Ternes}},
	  \bibinfo {author} {\bibfnamefont {S.}~\bibnamefont {Loth}}, \bibinfo {author}
	  {\bibfnamefont {C.~P.}\ \bibnamefont {Lutz}}, \bibinfo {author}
	  {\bibfnamefont {C.~F.}\ \bibnamefont {Hirjibehedin}}, \ and\ \bibinfo
	  {author} {\bibfnamefont {A.~J.}\ \bibnamefont {Heinrich}},\ }\href {\doibase
	  10.1103/PhysRevLett.103.107203} {\bibfield  {journal} {\bibinfo  {journal}
	  {Physical Review Letters}\ }\textbf {\bibinfo {volume} {103}},\ \bibinfo
	  {pages} {107203} (\bibinfo {year} {2009})}\BibitemShut {NoStop}%
	\bibitem [{\citenamefont {Br{\"{u}}ne}\ \emph {et~al.}(2012)\citenamefont
	  {Br{\"{u}}ne}, \citenamefont {Roth}, \citenamefont {Buhmann}, \citenamefont
	  {Hankiewicz}, \citenamefont {Molenkamp}, \citenamefont {Maciejko},
	  \citenamefont {Qi},\ and\ \citenamefont {Zhang}}]{Brune2012}%
	  \BibitemOpen
	  \bibfield  {author} {\bibinfo {author} {\bibfnamefont {C.}~\bibnamefont
	  {Br{\"{u}}ne}}, \bibinfo {author} {\bibfnamefont {A.}~\bibnamefont {Roth}},
	  \bibinfo {author} {\bibfnamefont {H.}~\bibnamefont {Buhmann}}, \bibinfo
	  {author} {\bibfnamefont {E.~M.}\ \bibnamefont {Hankiewicz}}, \bibinfo
	  {author} {\bibfnamefont {L.~W.}\ \bibnamefont {Molenkamp}}, \bibinfo {author}
	  {\bibfnamefont {J.}~\bibnamefont {Maciejko}}, \bibinfo {author}
	  {\bibfnamefont {X.-L.}\ \bibnamefont {Qi}}, \ and\ \bibinfo {author}
	  {\bibfnamefont {S.-C.}\ \bibnamefont {Zhang}},\ }\href {\doibase
	  10.1038/nphys2322} {\bibfield  {journal} {\bibinfo  {journal} {Nature
	  Physics}\ }\textbf {\bibinfo {volume} {8}},\ \bibinfo {pages} {486} (\bibinfo
	  {year} {2012})}\BibitemShut {NoStop}%
	\end{thebibliography}
	
	
	%
	
	\onecolumngrid
	\appendix
	\newpage
	
\section*{Supplemental Information}
\setlength{\belowcaptionskip}{-5mm}

	\setcounter{figure}{0}
	\renewcommand{\thefigure}{S\arabic{figure}}
	
	Details and comments relevant to the presented supplemental figures are given in the respective figure captions.
	\begin{figure*}[ht!]
		\center{\includegraphics{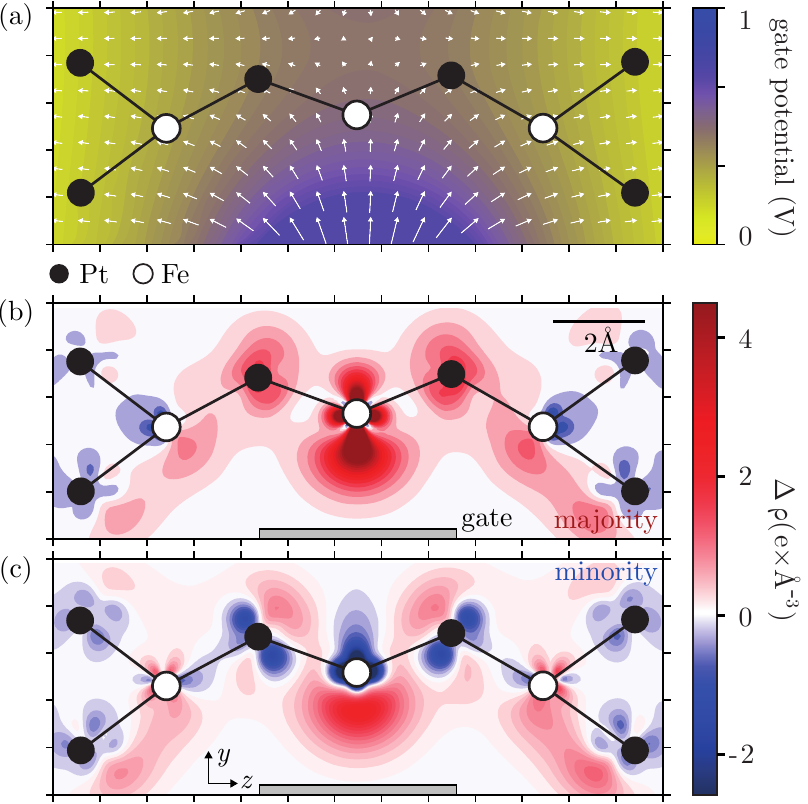}}
		\caption{Electrostatic potential and spin-dependent charge redistribution invoked by a gate electrode in the proximity of a zig-zag shaped Fe-Pt nano-contact (similar to Fig.~\ref{fig:pot_chredist}, where the same quantities for the linear configuration of the nano-contact are shown). (a) Electrostatic potential (color-coded) created by a gate electrode charged to 1\hole and the ensuing electric field (arrows, length proportional to the field strength). Charge redistribution $\Delta \rho$ of majority (b) and minority (c) electrons due to the presence of a biased gate electrode charged to 1\hole. Here $\Delta \rho_\sigma = \rho_\sigma (V_{gate}=V_{1\hole}) - \rho_\sigma (V_{gate}=0)$, where $\sigma$ denotes one of the two spin channels. The similarity of charge redistribution patterns to the case of a linear contact support our claim to the generality of the conclusions made in the main text of the manuscript.}
		\label{fig:s:pot_chredist}
	\end{figure*}
	\begin{figure*}[ht!]
		\center{\includegraphics{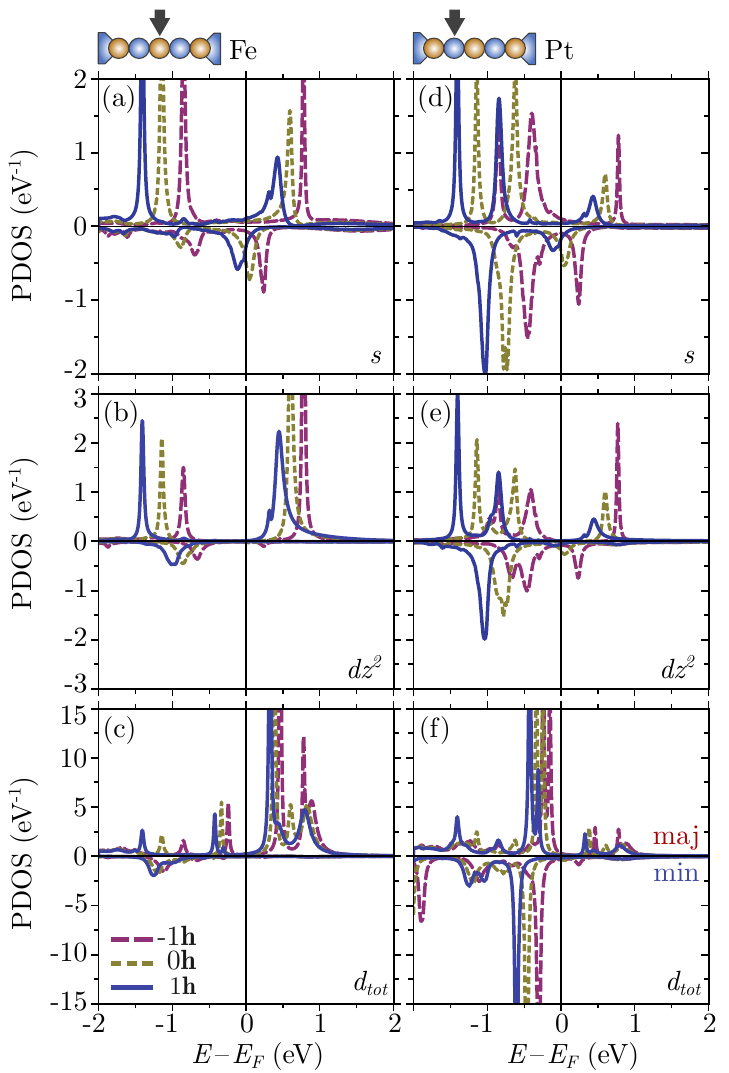}}
		\caption{Projected $s$ (a,d), $d_{z^2}$ (b,e) and $d_{tot}$ (c,d) densities of states at the central Fe (a-c) and the neighboring Pt (d-f) atoms of a linear Fe-Pt nano-contact with $d=4.75\AA$ at different gate electrode charges -- a broader view of the PDOS plotted in Fig.~\ref{fig:pdos_trc}. With changing gate charge the ``Stoner'' splitting between the coerrsponding majority and minority peaks decreases in accordance with the slight reduction of the magnetic moment of the atoms. The reduction of the splitting is acompanied with a general shift of the $d$ and $s$ bands (levels) of the cental Fe and the neigboring Pt atoms causing, in conjunction with the Stoner splitting reduction, the shift of the transmission peak across the Fermi lever as discussed in the main text of the manuscript.}
		\label{fig:s:pdos_wide}
	\end{figure*}
	\begin{figure*}[ht!]
		\center{\includegraphics{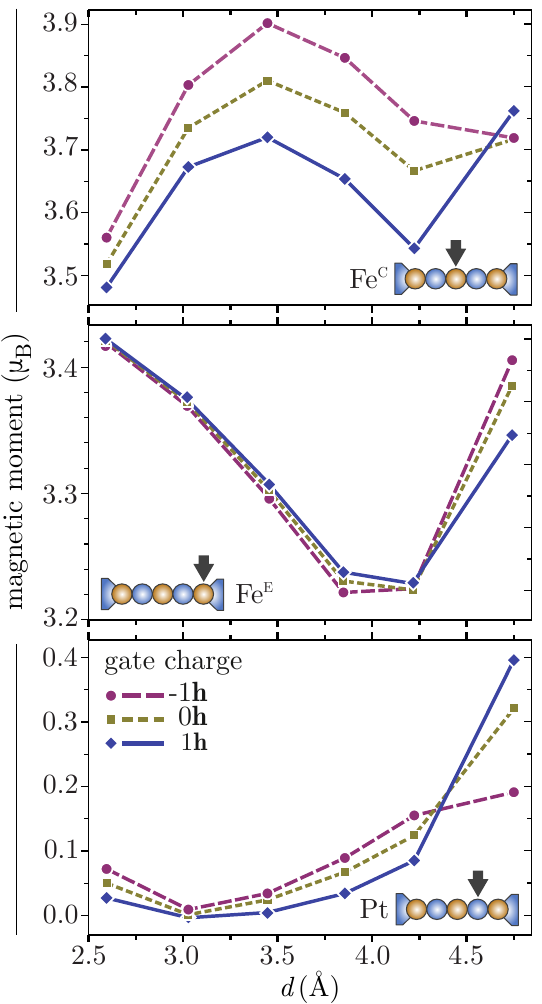}}
		\caption{Absolute value of the local magnetic moments of the central \chem{Fe^C} (a), edge \chem{Fe^E} (b) as well as \chem{Pt} (c) atoms of the nano-contact as a function of stretching parameter $d$ for different charge states of the gate electrode. The last point of each curve corresponds to the transition to a linear contact configuration which explains the qualitative change of the trend. The magnetization induced in the lead atoms is megligible and does not exceed $0.1~\mB$ for all $d$.}
		\label{fig:s:magmom}
	\end{figure*}
	\begin{figure*}[ht!]
		\center{\includegraphics{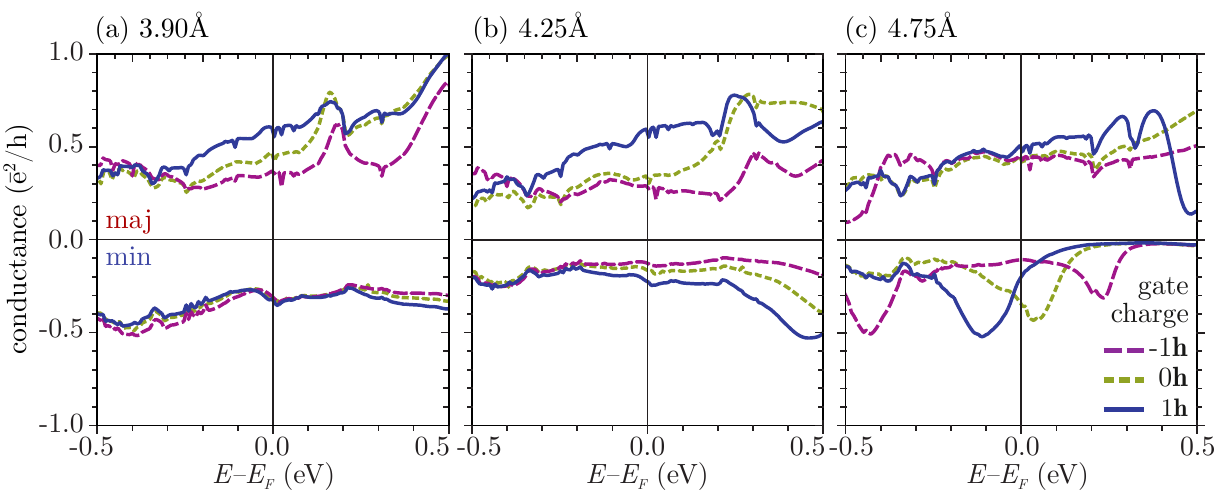}}
		\caption{Transmission probability of the Pt-Fe nano-contact for different distances $d$ between electrodes ($d=3.90$, $4.25$, $4.75\AA$ in panels (a), (b) and (c), respectively). Arrows denote spin orientations. The figure shows that the confined (or quantum well) states in the junction exist and are susceptible to the gate potential at different contact stretching states, in a way similar to the one described for the contact at $d=4.75\AA$ in the main text. This mechanism, however, does not always lead to a possibility of bias-induced conductance spin-polarization switching as the quantum well states' location with respect to the Fermi energy varies greatly depending on the contact geometry.}
		\label{fig:s:trc_dist}
	\end{figure*}
	\begin{figure*}[ht!]
		\center{\includegraphics{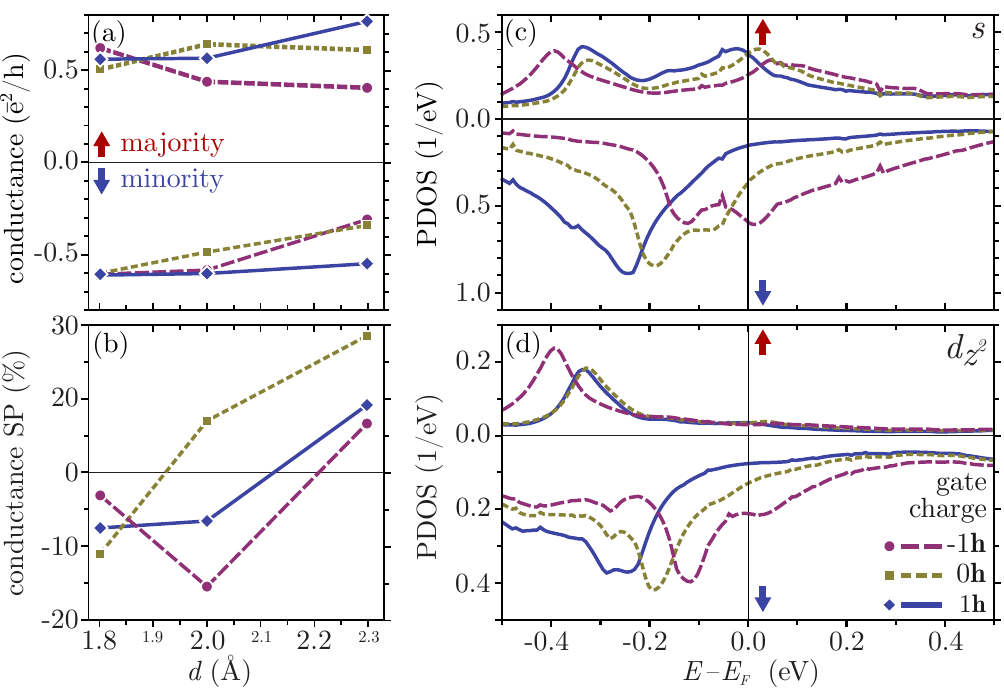}}
		\caption{Conductance (a) and its spin-polarization (b) of a Rh-Fe nano-contact for different distances $d$ and different gate charges. The behavior of spin-polarization of conductance shows that also in this system a switching of the spin-polarization of transport can be induced by a gate bias at short and intermediate inter-lead distances, when the contact is in a zig-zag geometric state. Projected density of $s$ (c) and $d_{z^2}$ (d) states of the Rh atom reveal the same physical mechanism responsible for the gate-induced alteration of the conductance spin polarization as discussed in the main paper of the paper for Pt-Fe nano-contacts. The quantum well states are observed in both majority and minority channels, yet their position is shifted to lower energies, showing that while the physics of the process is universal, particular realizations of the nanojunction in terms of chemical composition and its structural properties have to be chosen with care when considering potential technological applications of the bias-controlled spin-filtering phenomenon.}
		\label{fig:s:rhfe}
	\end{figure*}
	\begin{figure*}[ht!]
		\center{\includegraphics{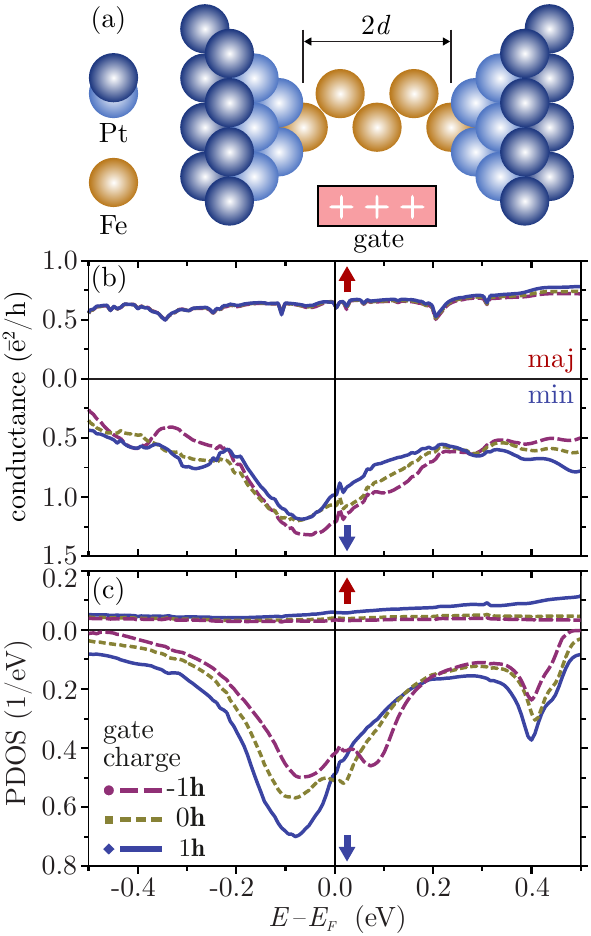}}
		\caption{Sketch of an all-Fe nano-contact (a) consisting of 5 Fe atoms suspended between Pt leads shaped as truncated pyramids. The system is similar geometrically to the alloyed contacts described in the main part of the paper, however the magnetic coupling between the atoms of the Fe chain is, contrary to the alloyed case, predominantly ferromagnetic, making the system more robust magnetically but less susceptible to manipulation. The spin-resolved conductance (b) and projected density of $p$-states (b) of the pure-Fe contact described above for different gate biases shows that the atomic like peak in the minority density of states close to the Fermi energy results in high spin-polarization of conductance and this robust spin-filtering quality of the contact is also susceptible to gating, yet the impact the gate biases of a realistic magnitude is not strong enough to allow for a switching of the spin-filtering properties of the junction.}
		\label{fig:s:pure_fe}
	\end{figure*}
\end{document}